\documentclass[preprint2,tighten]{aastex6}
\pdfoutput=1 
\usepackage{amsmath,amstext}
\usepackage[T1]{fontenc}
\usepackage{apjfonts} 
\usepackage{appendix}
\usepackage[figure,figure*]{hypcap}
\usepackage{color}
\usepackage[figuresright]{rotating}
\shorttitle{}
\shortauthors{Peters et al.}

\begin{document}

\title{The Hubble Space Telescope Advanced Camera for Surveys Emission Line Survey of Andromeda. I: Classical Be Stars}

\author{
M. Peters\altaffilmark{1}, 
J.P. Wisniewski\altaffilmark{1},
B.F. Williams\altaffilmark{3},
J.R. Lomax\altaffilmark{7},
Y. Choi\altaffilmark{2},
M. Durbin\altaffilmark{3},
L.C. Johnson\altaffilmark{5},
A.R. Lewis\altaffilmark{6},
J. Lutz\altaffilmark{3},
T.A.A. Sigut\altaffilmark{4},
A. Wallach\altaffilmark{3},
J.J. Dalcanton\altaffilmark{3}
}

\altaffiltext{1}{Homer L. Dodge Department of Physics, University of Oklahoma, 440 W. Brooks St., Norman, OK 73071, USA}
\altaffiltext{2}{Steward Observatory, University of Arizona, 933 North Cherry Avenue, Tucson AZ 85721}
\altaffiltext{3}{Department of Astronomy, University of Washington, Box 351580, Seattle, WA 98195, USA}
\altaffiltext{4}{Department of Physics and Astronomy, University of Western Ontario, London, ON Canada, N6A 3K7}
\altaffiltext{5}{Department of Physics and Astronomy, Northwestern University, 2145 Sheridan Road, Evanston, IL 60208, USA}
\altaffiltext{6}{Center for Cosmology and AstroParticle Physics, The Ohio State University, Columbus, OH 43210, USA; Department of Astronomy, The Ohio State University, 140 West 18th Avenue, Columbus, OH 43210, USA}
\altaffiltext{7}{United States Naval Academy, Physics Department, 572 C Holloway Rd, Annapolis, MD, 21402 USA}

\begin{abstract}

We present results from a 2-epoch HST H$\alpha$ emission line survey of the Andromeda Galaxy that overlaps the footprint of the Panchromatic Hubble Andromeda Treasury (PHAT) survey. We find 552 (542) classical Be stars and 8429 (8556) normal B-type stars in epoch \# 1 (epoch \# 2), yielding an overall fractional Be content of 6.15\% $\pm$0.26\% (5.96\% $\pm$0.25\%). The fractional Be content decreased with spectral sub-type from $\sim$23.6\% $\pm$2.0\% ($\sim$23.9\% $\pm$2.0\%) for B0-type stars to $\sim$3.1\% $\pm$0.34\% ($\sim$3.4\% $\pm$0.35\%) for B8-type stars in epoch \# 1 (epoch \# 2). We observe a clear population of cluster Be stars at early fractional main sequence lifetimes, indicating that a subset of Be stars emerge onto the ZAMS as rapid rotators. Be stars are 2.8x rarer in M31 for the earliest sub-types compared to the SMC, confirming that the fractional Be content decreases in significantly more metal rich environments (like the Milky Way and M31).   However, M31 does not follow a clear trend of Be fraction decreasing with metallicity compared to the Milky Way, which may reflect that the Be phenomenon is enhanced with evolutionary age. The rate of disk-loss or disk-regeneration episodes we observe, 22\% $\pm$ 2\% yr$^{-1}$, is similar to that observed for seven other Galactic clusters reported in the literature, assuming these latter transient fractions scale by a linear rate.  The similar number of disk-loss events (57) as disk-renewal events (43) was unexpected since disk dissipation time-scales can be $\sim$2x the typical time-scales for disk build-up phases.

\end{abstract}
\section{Introduction}

Classical B-type emission line (Be) stars are non-supergiant stars whose spectrum exhibit or have exhibited H$\alpha$ emission lines \citep{Jaschek1981}, interpreted as arising from an optically-thin gaseous circumstellar disk \citep{struve}.  A defining property of classical Be stars is their rapid rotation rate, which is estimated to be 
upwards of 75 percent of the critical rotation rate \citep{porter,Rivinius2013}.  The velocity 
structure of gas in these disks is consistent with Keplerian motion \citep{Hummel,mei,wheel} 
and their behavior is well explained by the viscous decretion disk paradigm of \citet{Lee}.

A long-standing challenge in the study of classical Be stars has been identifying the 
mechanism(s) responsible for launching material from the stellar photosphere, while remaining 
consistent with both the observed velocity structure and the ability of systems to switch between periods 
in which they exhibit a disk and quiescent periods where all evidence for the presence of a disk disappears.  As summarized in \citet{Rivinius2013}, non-radial pulsations have been 
observed for a growing number of Be stars (e.g. \citealt{neiner2002,Rivinius2003,Baade2016,Bartz2017}), and could be one mechanism that contributes to the ejection of material into Be disks.  \citet{baade16,baade18} have suggested that so-called ``Stefl'' frequency of variations might trace star-to-disk mass transfer.  Internal gravitational wave pulsation modes might also help transport angular momentum from the core to stellar surface \citep{rogers2013}; \citet{neiner2012} has shown that stochastically driven g-modes might operate in a subset of Be stars, in addition to the more commonly observed Kappa driven p- and g-modes.  The role of magnetic fields
has been considered \citep{Cassinelli2002,ud} but observational studies have failed to find evidence of large-scale fields in classical Be stars, despite their observed presence around non-Be B-type stars \citep{wade}. Recent theoretical magnetohydrodynamical simulations of Be disks have corroborated this by showing that field strengths of only 10G can disrupt Keplerian circumstellar disks \citep{ud2018}.


Modern stellar evolution models that account for the role of rotation have demonstrated how the metallicity of stars can affect the internal evolution of their angular momentum on the main sequence \citep{mae2000,mae2001}.  It is therefore perhaps not surprising that observations of massive stars in the Galaxy, LMC, and SMC have shown that the prevalence of classical Be star-disk systems, which are characterized by their rapid rotation, increases with decreasing metallicity \citep{Grebel92,Maeder1999,McSwain2005,Wisniewski2006,Martayan2006,martayan2010,iq2013}.  If the fractional increase in the presence of Be stars with decreasing metallicity is indeed caused by stars in low metallicity environments having higher stellar rotational velocities \citep{mar07,mae2000,mae2001}, then one would predict that the prevalence of the Be phenomenon in M31, which is on average 1.5x Solar metallicity \citep{Clayton2015}, would be even less common than in our Galaxy. 

Numerous observational and theoretical studies have explored
the role that evolutionary age plays in the Be phenomenon.  Stellar evolution models that incorporate the effects of rotation predict that mass loss can dramatically reduce $\Omega$/$\Omega_{crit}$ throughout the main sequence evolution of O- and early B-type stars in Galactic-metallicity environments \citep{meynet2000}, but $\Omega$/$\Omega_{crit}$ can increase, especially near the TAMS, in lower metallicity environments like the SMC \citep{mae2001}.  Given the critical role of rapid rotation in Be disk formation, theory thus suggests that in addition to metallicity, evolutionary age could also play an important role in Be disk formation.  

Observationally, it is clear that at least some classical Be stars must emerge on the ZAMS at near critical rotation rates (e.g. \citealt{wisniewski2007}).  Moreover, other studies have suggested that disk formation preferrentially happens in the second-half of a star's main sequence lifetime \citep{fabregat2000,martayan2010}, due to internal evolution of a star's angular momentum \citep{meynet2000,mae2001} and/or spin-up by companions \citep{McSwain2005}.  Further observations that trace the frequency of the Be phenomenon with main sequence lifetime, stellar mass, and metallicity are needed to discern what drives changes in the internal rotational behavior of B-type stars and whether the effects of metallicity and age can be disentangled.

In this paper, we present the first deep, wide-area, space-based H$\alpha$ emission line survey of a portion of M31 that overlaps with the HST-PHAT survey.  We focus our analysis on the classical Be star population in M31.  We describe the acquisition and reduction of the data in Section \ref{sec:Observation}.  We outline our method of identifying classical Be stars in our sample, present the fractional Be content of our sample as a function of spectral type, and detail differences observed in our Be disk population between our two epochs of observations in Section \ref{sec:Analysis}.  Finally, we then discuss the implications of the multi-epoch behavior of our data, the dependence of the Be phenomenon on age and metallicity, and the frequency of the Be phenomenon with spectral type in Section \ref{sec:Discussion}.

\section{Observations and Data Reduction} \label{sec:Observation}
\subsection{Observational Design}\label{sec:ObsDes}
\begin{figure}[htp] \label{Footprints}
\includegraphics[width=\columnwidth]{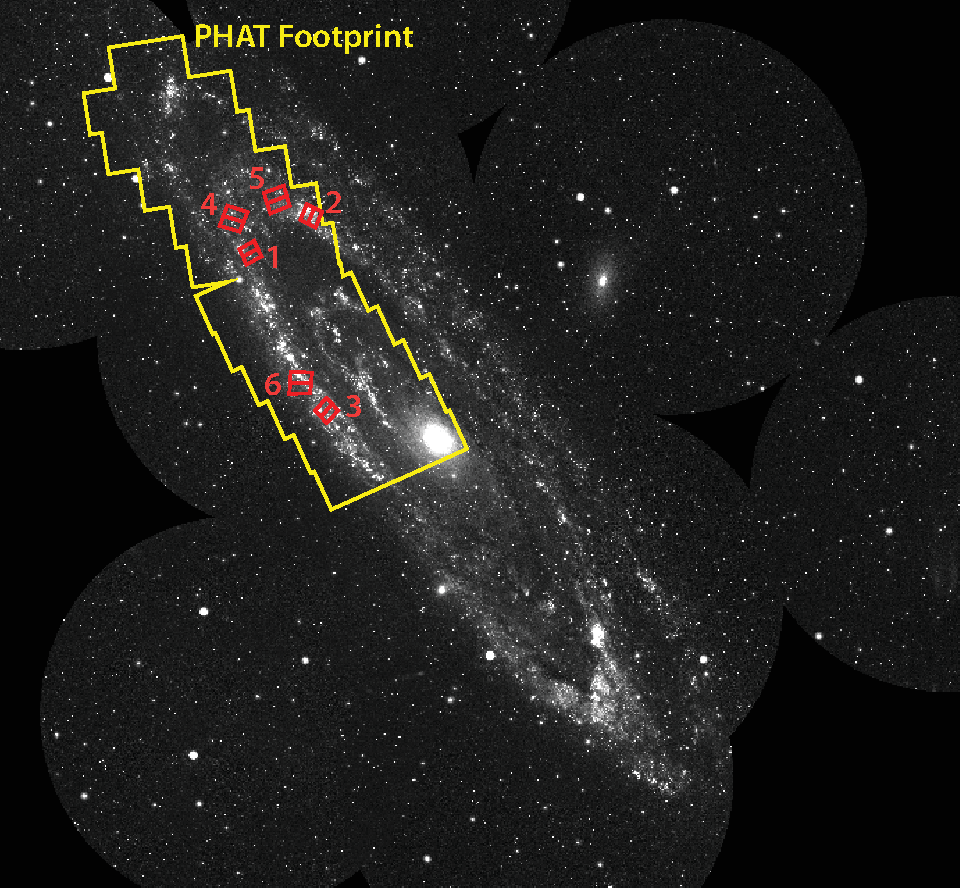}
\caption{The footprint of our HST emission line survey pointings of M31 are depicted in \textbf{red}, and labeled by pointing number compiled in Table \ref{tbl:hstdetails}.  Also shown here in \textbf{yellow} is the overlay of the full PHAT survey \citep{dalcanton2012}.}
\end{figure}

We observed M31 with the Hubble Space Telescope (HST) in the 12-orbit GO-13857 program (Dalcanton PI).  Pure-parallel observations 
were made with the ACS/WFC and WFC3/UVIS camera at three distinct 
spatial locations that coincided with regions of the galaxy previously 
observed via the HST/PHAT survey \citep{dalcanton2012}, yielding 6 distinct pointings across M31 (Figure \ref{Footprints}).  We obtained a second epoch of observations of these same 6 pointings approximately 1 year later, when HST was able to achieve the same view of these pure-parallel pointings.    
Observations were made in both cameras using the \texttt{F658N} narrow-band filters and using the \texttt{F625W} broad-band filter.  A summary of our basic observational design is provided in Table \ref{tbl:hstdetails}.  As discussed and explained in Section \ref{allBe}, although we present the emission line sample derived from both the ACS/WFC and WFC3/UVIS data, we restrict our Be star statistical analysis to the ACS/WFC data to avoid biasing our results due to the slightly different central wavelengths of the \texttt{F658N} filter between the two cameras.

\begin{table*}
\begin{center}
\caption{Summary of HST Observations}\label{tbl:hstdetails}
\tablewidth{0pc}
\tablecolumns{3}
\begin{tabular}{ccccccc}
\\
\tableline\tableline
Epoch \# & Pointing & Date of Obs & Camera & Coordinates & \texttt{F625W} Expos Time & \texttt{F658N} Expos Time \\
\nodata & \nodata & \nodata & \nodata & \nodata & sec. & sec. \\
\tableline
1 & 1 & 2014 Oct 28 & WFC3 & 00:45:23.38, +41:46:12.5 & 750 & 3850 \\
2 & 1 & 2015 Oct 5 & WFC3 & 00:45:23.38, +41:46:12.5 & 750 & 3850 \\
1 & 2 & 2014 Oct 1 & WFC3 & 00:44:31.02, +41:52:08.5 & 750 & 3850 \\
2 & 2 & 2015 Aug 29 & WFC3 & 00:44:31.02, +41:52:08.5 & 750 & 3850 \\
1 & 3 & 2014 Sep 30 & WFC3 & 00:44:18.05, +41:20:53.7 & 750 & 3850 \\
2 & 3 & 2015 Sep 25 & WFC3 & 00:44:18.05, +41:20:53.7 & 750 & 3850 \\
1 & 4 & 2014 Oct 28 & ACS/WFC & 00:45:38.48, +41:51:27.3 & 750 & 3850 \\
2 & 4 & 2015 Oct 5 & ACS/WFC & 00:45:38.49, +41:51:27.3 & 750 & 3850 \\
1 & 5 & 2014 Oct 1 & ACS/WFC & 00:45:00.29, +41:54:32.6 & 750 & 3850 \\
2 & 5 & 2015 Aug 29 & ACS/WFC & 00:45:00.29, +41:54:32.6 & 750 & 3850 \\
1 & 6 & 2014 Sep 30 & ACS/WFC & 00:44:40.57, +41:25:05.3 & 750 & 3850 \\
2 & 6 & 2015 Sep 25 & ACS/WFC & 00:44:40.57, +41:25:05.3 & 750 & 3850 \\

\tableline
\end{tabular}
\tablecomments{Summary of HST observations of M31 from GO-13857.}
\end{center}
\end{table*}

\subsection{Data Reduction} We utilized the same methodology and pipeline for extracting photometry from our data as used in the PHAT survey, as described in \citet{dalcanton2012} and \cite{williams2014}.  In brief, the images from each epoch of observing (2014 and 2015) for each of the six fields were separately processed through our Cloud-based photometry pipeline \citep{williams2018}, which uses the \texttt{DOLPHOT} photometry package \citep{dolphot2000}
to force a fit of the pre-computed point spread function on every resolved star in the image stack of greater than 2.5-$\sigma$ significance above the expected background level.  The photometry from these stars was then aligned and matched to the full 117 million source catalog from the full PHAT survey as presented in \citet{williams2014} to allow us to compare the stars' narrow band properties to their 6-band spectral energy distributions.  Overall, the photometric reductions were of comparable quality to those of the PHAT survey, except that we found a systematic uncertainty between epochs.   Our analysis showed that our broadband exposures covered a short enough portion of the orbit that the thermal expansion and contraction of the optics caused a significant offset between the orbit-averaged PSF used by \texttt{DOLPHOT} and that of the \texttt{F625W} images across epochs, which introduced a systematic error that we further discuss in Appendix A.    

Next, our source catalog was enhanced by adding the output of the Bayesian Extinction And Stellar Tool (\texttt{BEAST}; \citealt{gordon2016}) on the PHAT catalog data, which added T$_{eff}$, log g, and age information for each source. \citet{gordon2016} developed the \texttt{BEAST} to model the observed broad spectral energy distribution (SED) of an individual star by considering the full observational uncertainties. The \texttt{BEAST} primarily returns the posterior probability distributions for the intrinsic stellar properties of individual stars (e.g., stellar temperature, age, surface gravity) and their line of sight dust information (dust column density, grain size distribution, composition). The \texttt{BEAST} has been successfully used for many multi-band HST programs such as PHAT (PI: Dalcanton), SMIDGE (PI: Sandstrom), METAL (PI: Roman-Duval), and the Local Volume UV survey (PI: Gilbert). Specifically, \citet{gordon2016} ran the \texttt{BEAST} for all PHAT stars that were detected in \texttt{F475W} and at least 3 more bands (among \texttt{F225W}, \texttt{F336W}, \texttt{F814W}, \texttt{F110W}, \texttt{F160W}) with resolutions of 0.15 in log age, 0.15 in Av, 0.25 in fA and Z=0.004, 0.008, 0.019, and 0.03. For this study, we cross match this PHAT-\texttt{BEAST} catalog with our classical Be star candidates, and use the \texttt{BEAST} measurements to better categorize and understand Be stars in M31. Finally, we corrected our data for focus-induced offsets, as described in Appendix \ref{oops}.

\section{Analysis}\label{sec:Analysis}
\subsection{Defining the Sample}
\begin{figure}[htp]
\includegraphics[width=\columnwidth]{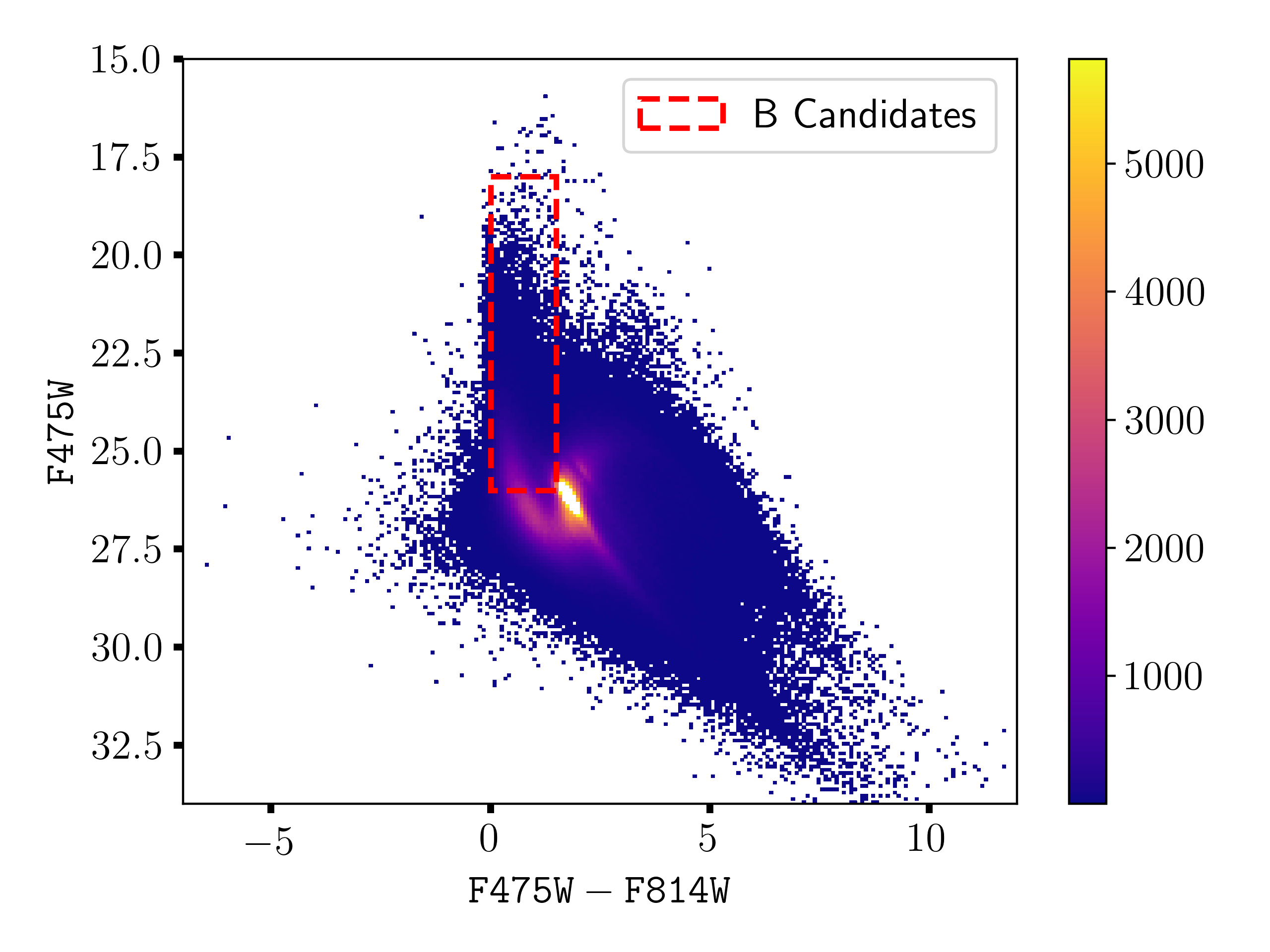}
\caption{A color magnitude diagram of all $\sim$2 million point sources detected via our program, without excluding any sources based on their intrinsic properties or observed errors.}
\label{fullcmd}
\end{figure}

The initial reduced data set contained approximately two million stars 
in each epoch (Figure \ref{fullcmd}). As this paper focuses solely on the analysis of 
the classical Be population of the survey, we applied several criteria 
to isolate the population of main sequence B-type stars with high quality data from the larger survey database.  Our basic process was to utilize the value-added nature of \texttt{BEAST}-derived fundamental stellar parameters to isolate the population of main sequence B-type stars in our sample, and then use traditional color magnitude diagrams (CMDs) to confirm the robustness of these inferred stellar parameters.

We began our sample definition by requiring sources in the new observations to have a signal-to-noise ratio (SNR) of $>$10 in the \texttt{F475W}, \texttt{F625W}, \texttt{F658N}, and \texttt{F814W} filters, which are the primary filters that we used in our subsequent 
analysis.  Next, we required sources have \texttt{BEAST}-derived 50th-percentile effective temperatures that resided within 
the range expected for B-type stars, 4.00 $<$ Log(T$_{eff}$) $<$ 4.51 \citep{underhill}. We also required that sources have a \texttt{BEAST}-derived 50th-percentile surface gravity, inclusive of +$1\sigma_{err}$ computed from the 84th-percentile surface gravity, of Log(g)$>$ 3.75, to isolate main sequence stars and exclude giants and supergiants \citep{Castelli2004}.  

\begin{figure}[htp]
\includegraphics[width=\columnwidth]{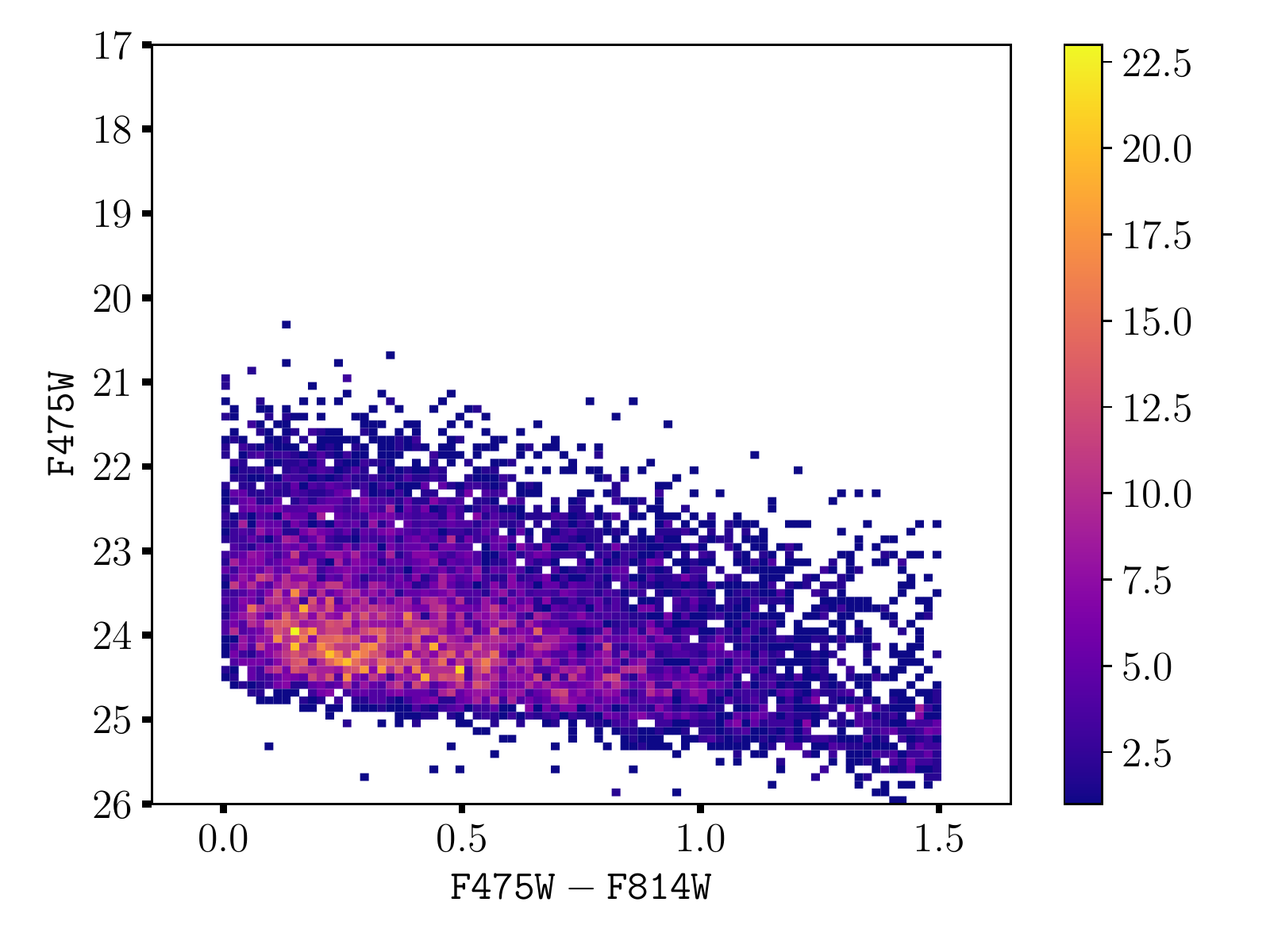}
\caption{A color magnitude diagram of all point sources that remained after the application of the data quality criteria.}
\label{cutscmd}
\end{figure}

We inspected the distribution of these selected B-star candidates on CMDs, to determine whether uncertainties associated with \texttt{BEAST}-derived stellar parameters were introducing large populations of likely false-positive B-type main sequence stars.  We found that adopting a more conservative surface gravity criteria (Log(g)$_{16}$ $>$ 3.75, where Log(g)$_{16}$ is the 16th percentile of the Log(g) posterior probability distribution) led to a 4x reduction in the number of sources, but these extra sources that were removed occupied the same region on the CMD as our higher confidence main sequence B stars. As such, we adopted our less conservative criteria (Log(g)$_{84}$ $>$ 3.75), which had the net effect of being most effective at excluding only the supergiants from our sample while retaining the main sequence stars. Based on these inspections, we included additional criteria to reject sources having (\texttt{F475W}$-$\texttt{F814W}) $<$ 0, which were likely planetary nebulae (see e.g. \citealt{v2014}), and also rejected objects having (\texttt{F475W}$-$\texttt{F814W}) $>$ 1.5, as these sources had \texttt{BEAST}-derived Log(g) values near 3.75 and were clearly isolated from the bulk population with larger Log(g) values, indicating they were likely post-main-sequence giants. Finally, we excluded a small number of sources in our catalog that were assigned the same RA/Dec coordinates by our matching routine between filters, due to source confusion between different survey tiles.  The final dataset for pointings 4-6 matching the aforementioned criteria contained 8,981 stars in epoch one and 9,098 in epoch two, and their CMD positions are plotted in Figure \ref{cutscmd}.

\subsection{Identifying Be stars}\label{identify}
Classical Be stars by definition are stars that exhibit or have exhibited H$\alpha$ emission, which is interpreted as originating from their circumstellar gas disks \citep{Jaschek81}.  As such, one technique commonly used to identify classical Be stars is 
the use of photometric 2-color diagrams (2-CDs) that utilize a narrow-band filter centered 
on H$\alpha$ and an associated filter that samples the nearby continuum region (see 
e.g. \citealt{Grebel92,Keller99,McSwain2005a,McSwain2005,Wisniewski2006}). We discuss some of the limitations of this diagnostic in more detail in Section \ref{limits}. We adopt a 
similar diagnostic approach here, using new data in the \texttt{F625W} and \texttt{F658N} HST filters 
from our GO-13857 program to diagnose the presence of emission or absorption in 
H$\alpha$, along with archival data in the \texttt{F475W} and \texttt{F814W} HST filters from the PHAT survey \citep{williams2014}.

We first computed synthetic photometry through the known properties of HST's optics and 
cameras for normal B-type main sequence stars spanning a range of B sub-types 
in the \texttt{F625W} and \texttt{F658N} filters.  We began by using \texttt{pysynphot} \citep{pysynphot2013} and the Castelli-Kurucz Atlas \citep{Castelli2004} to create synthetic stellar spectra for five sub-types B0, B1, B3, B5, and B8, adopting the stellar parameters listed in Table \ref{tbl:kurucz}, based on average values for each sub-type compiled in \citet{underhill}.  These spectra were created using 
\texttt{pysynphot}'s \texttt{Icat} function, which interpolates between the discrete models in the 
Castelli-Kurucz atlas.  We adopted a color excess of E(B$-$V) = 0.4, appropriate for moderately reddened 
stars in M31 \citep{Clayton2015}, along with R$_v$ = 3.1.  This adopted color excess is 
consistent with the average E(B$-$V) values computed for our sample of M31 B-type stars via the \texttt{BEAST} \citep{BEAST}.  We note that the adoption of a color excess appropriate for moderately 
reddened stars will lead to a more stringent color criterion for the identification 
of Be stars in our data, such that our Be star rates will be a lower limit.  We used the \texttt{Extinction} function to apply reddening to our spectra, yielding photospheric H$\alpha$ absorption line profiles that are representative of main sequence B-type stars (or classical Be stars when they are in a ``disk-less'' state).  We compile the equivalent widths of these H$\alpha$ absorption lines in Table \ref{tbl:kurucz}.


\begin{table*}
\begin{center}
\caption{Synthetic Spectra Properties}\label{tbl:kurucz}
\tablewidth{0pc}
\tablecolumns{7}
\begin{tabular}{ccccc}
\\
\tableline\tableline
Spectral Type & Temperature & Log(Gravity) & Log(Metallicity) & H$\alpha$ EW (\AA) \\
\tableline
B0V & 30000 & 3.90 & -1.7 & 2.83  \\
B1V & 25400 & 3.90 & -1.7 & 3.52   \\
B3V & 18700 & 3.94 & -1.7 & 4.67   \\
B5V & 15400 & 4.04 & -1.7 & 5.71  \\
B8V & 11900 & 4.04 & -1.7 & 7.52  \\
\tableline
\end{tabular}
\tablecomments{Synthetic spectra parameters were obtained from the recommended values in the Castelli-Kurucz Atlas; the metallicity adopted was 1.5x solar, as taken from \citet{Clayton2015}.} 
\end{center}
\end{table*}

The amount of H$\alpha$ emission that a classical Be star exhibits is to first 
order representative of the amount of disk material present within the stellocentric 
distances probed by H$\alpha$ \citep{Rivinius2013} at any epoch.
We next used the absorption spectra we created in the previous step to create emission-line spectra, representative of a classical Be star with an arbitrarily strong disk (e.g. having H$\alpha$ EWs as compiled in Table \ref{tbl:kurucz}, albeit in emission), and spectra whose photospheric absorption is merely filled in with emission, representative of a classical Be star with a more tenuous disk (e.g. having H$\alpha$ EWs of 0). We fit a Guassian to the H$\alpha$ line using \texttt{curve\_fit} from \texttt{scipy.optimize}, and then used this expression to create profiles with filled-in photospheric absorption and pure emission signatures. Finally, we computed the synthetic magnitudes in the \texttt{F625W} and \texttt{F658N} filters for the \texttt{ACS} and \texttt{WFC3} cameras for these spectra for each B sub-type to allow us to deduce which of the B-type stars in our sample were plausibly classical Be stars. 

\begin{figure}[htp]
\includegraphics[width=\columnwidth]{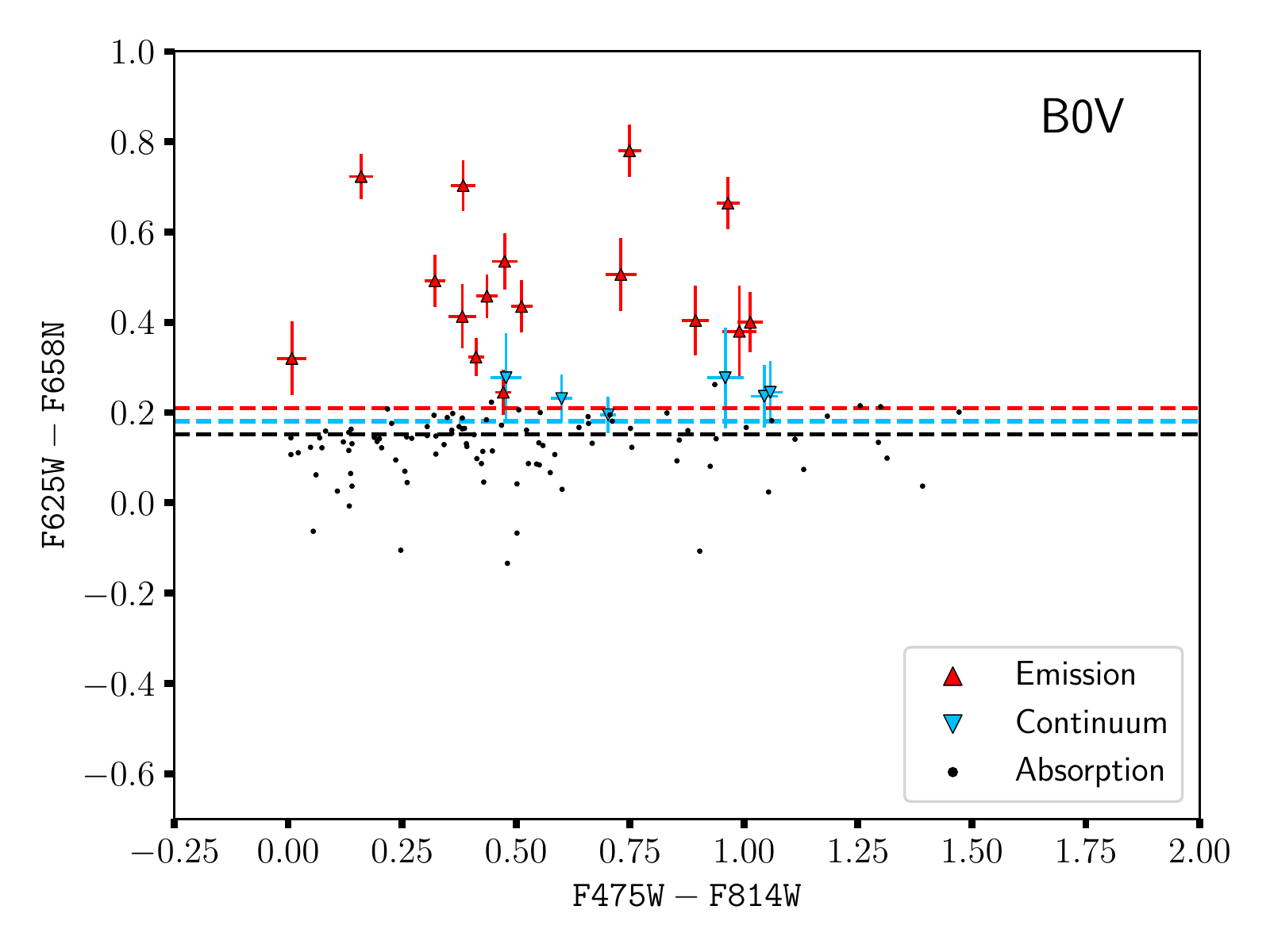}
\caption{A representative 2-CD for our B0V spectral type sources, from epoch 1, pointing 5. We classify objects as classical Be stars if their (\texttt{F625W}$-$\texttt{F658N}) colors, including 3$\sigma$ errors, are redder than the photospheric absorption level.  The horizontal dashed lines in these figures depict the colors expected for pure H$\alpha$ photospheric absorption (black), filled-in absorption (blue), and pure H$\alpha$ emission profiles as computed from our models. Note that the plotted error bars are 3$\sigma$ errors.} 
\label{2ColorB0V}
\end{figure}

\begin{figure}[htp]
\includegraphics[width=\columnwidth]{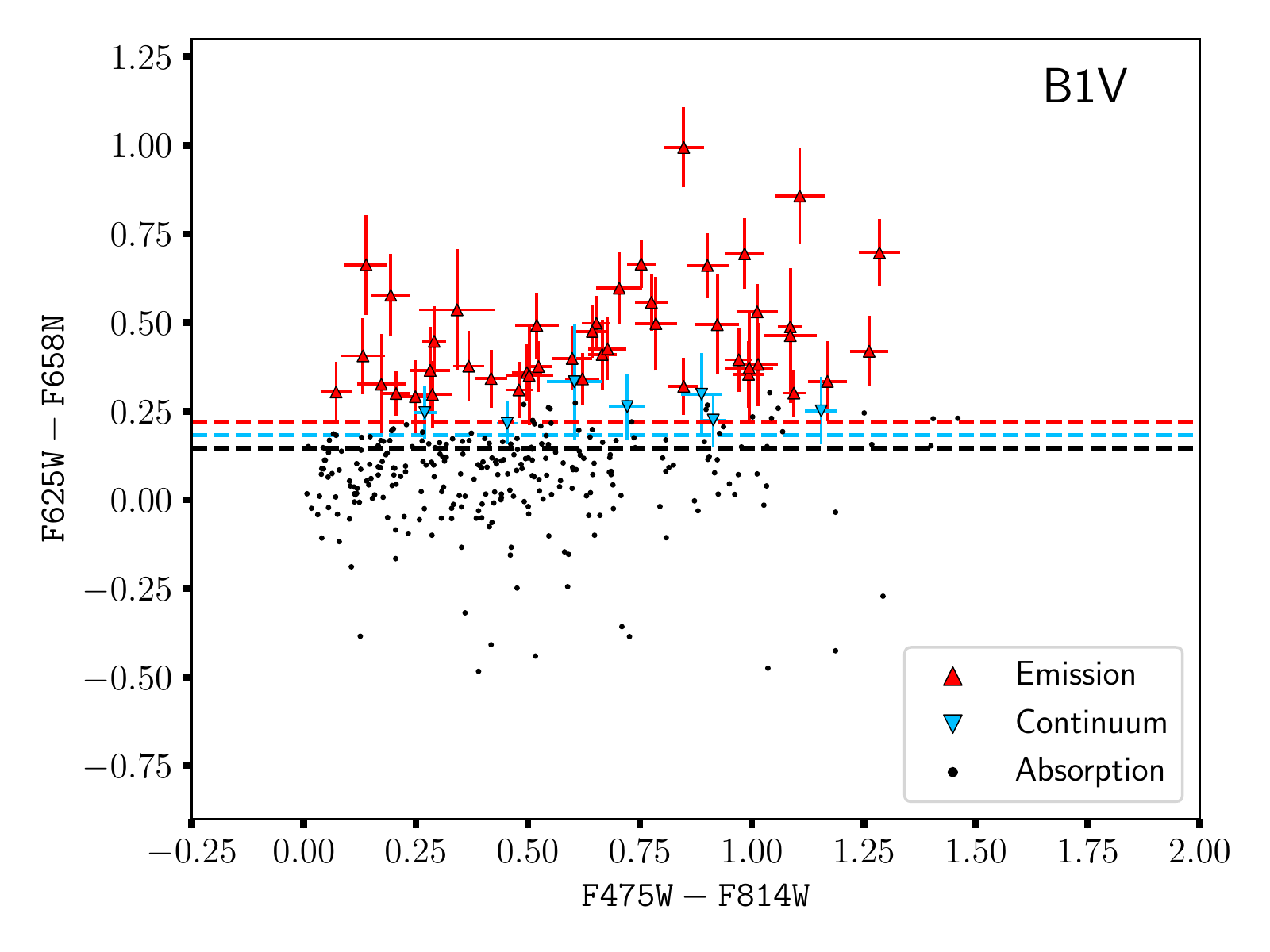}
\caption{A representative 2-CD for our B1V spectral type sources, from 1 epoch of a pointing.  The description of the data presentation is the same as in Figure \ref{2ColorB0V}.}
\label{2ColorB1V}
\end{figure}

\begin{figure}[htp]
\includegraphics[width=\columnwidth]{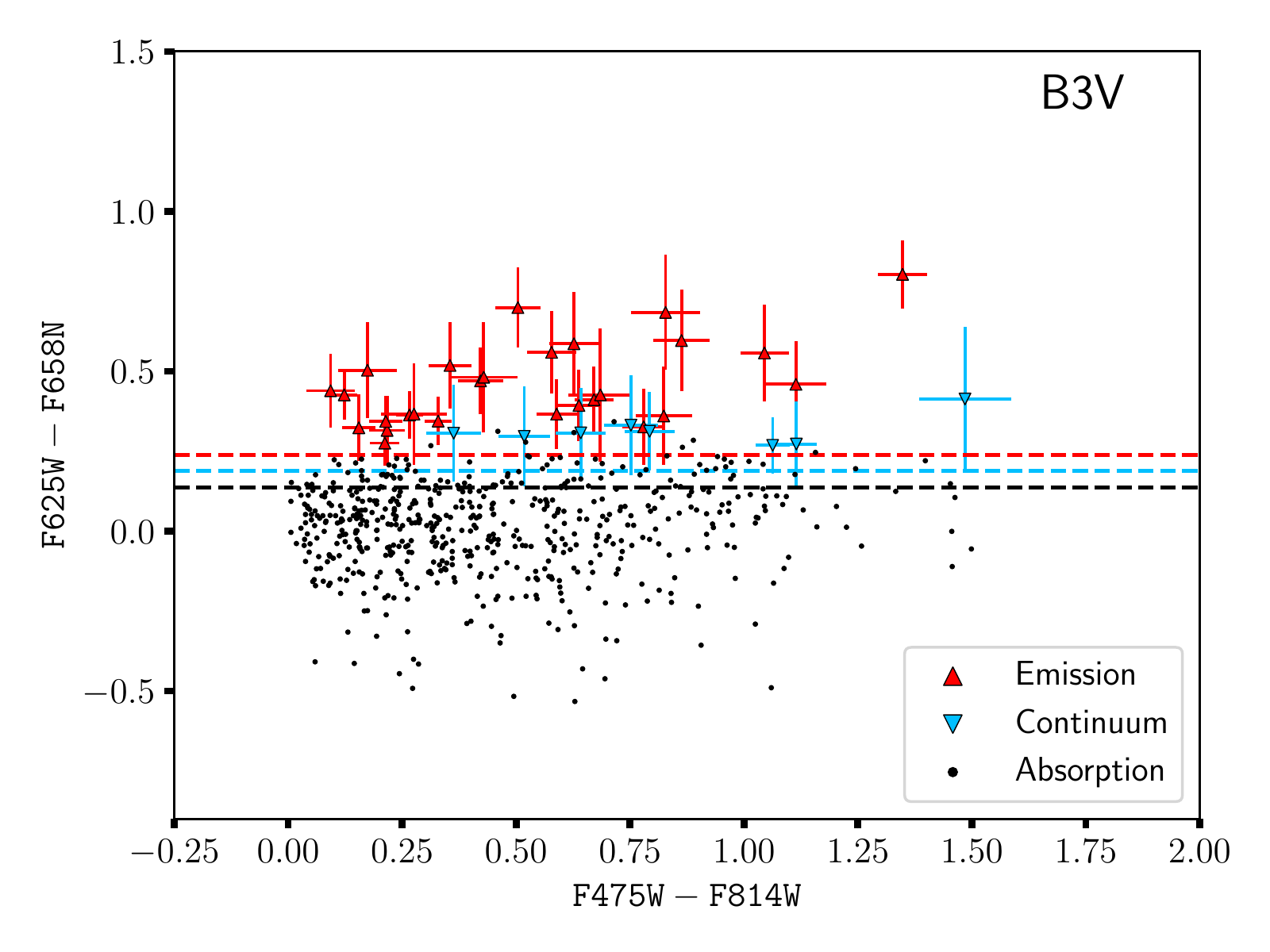}
\caption{A representative 2-CD for our B3V spectral type sources, from 1 epoch of a pointing.  The description of the data presentation is the same as in Figure \ref{2ColorB0V}.}
\label{2ColorB3V}
\end{figure}

\begin{figure}[htp]
\includegraphics[width=\columnwidth]{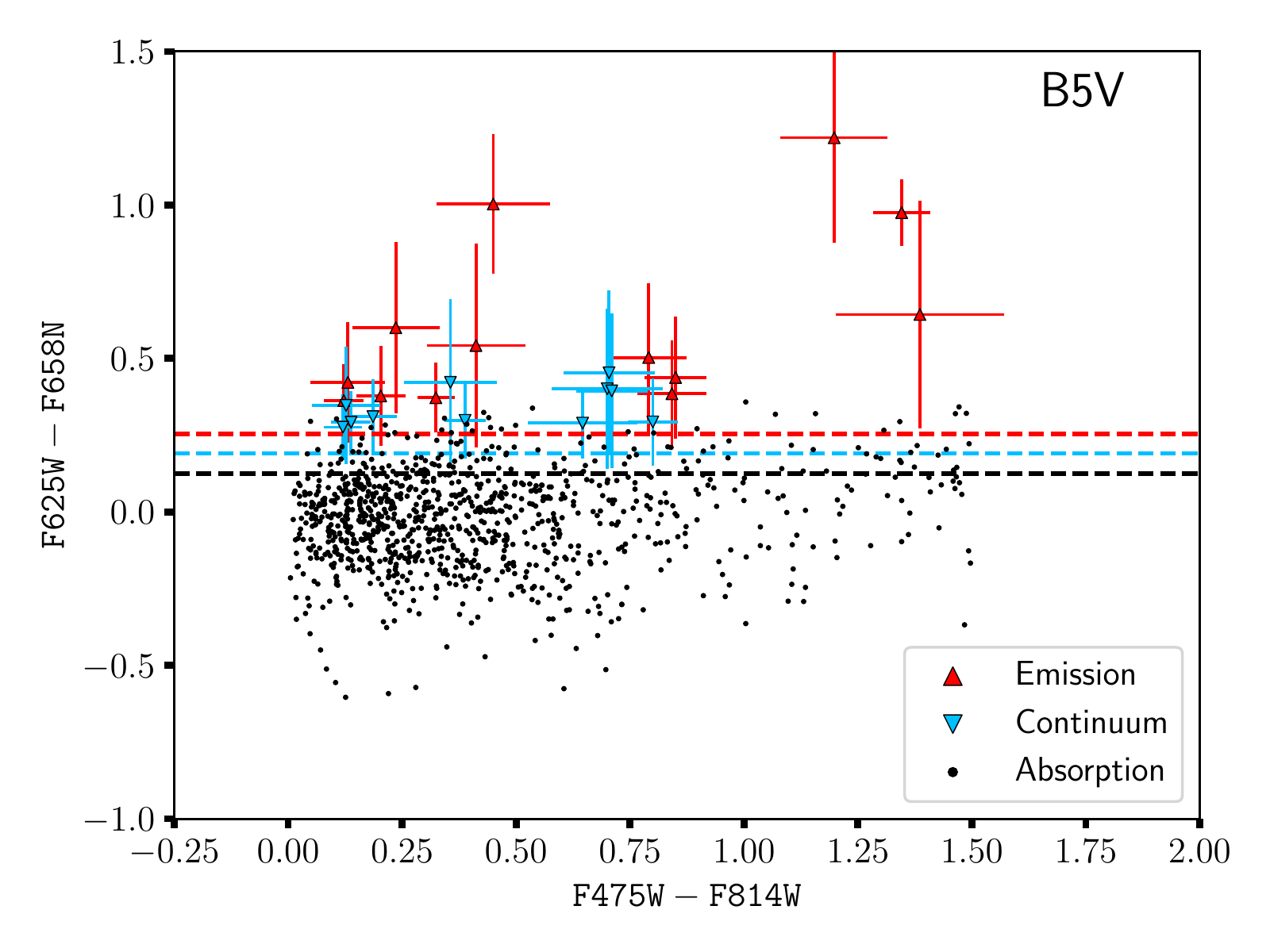}
\caption{A representative 2-CD for our B5V spectral type sources, from 1 epoch of a pointing.  The description of the data presentation is the same as in Figure \ref{2ColorB0V}.}
\label{2ColorB5V}
\end{figure}

\begin{figure}[htp]
\includegraphics[width=\columnwidth]{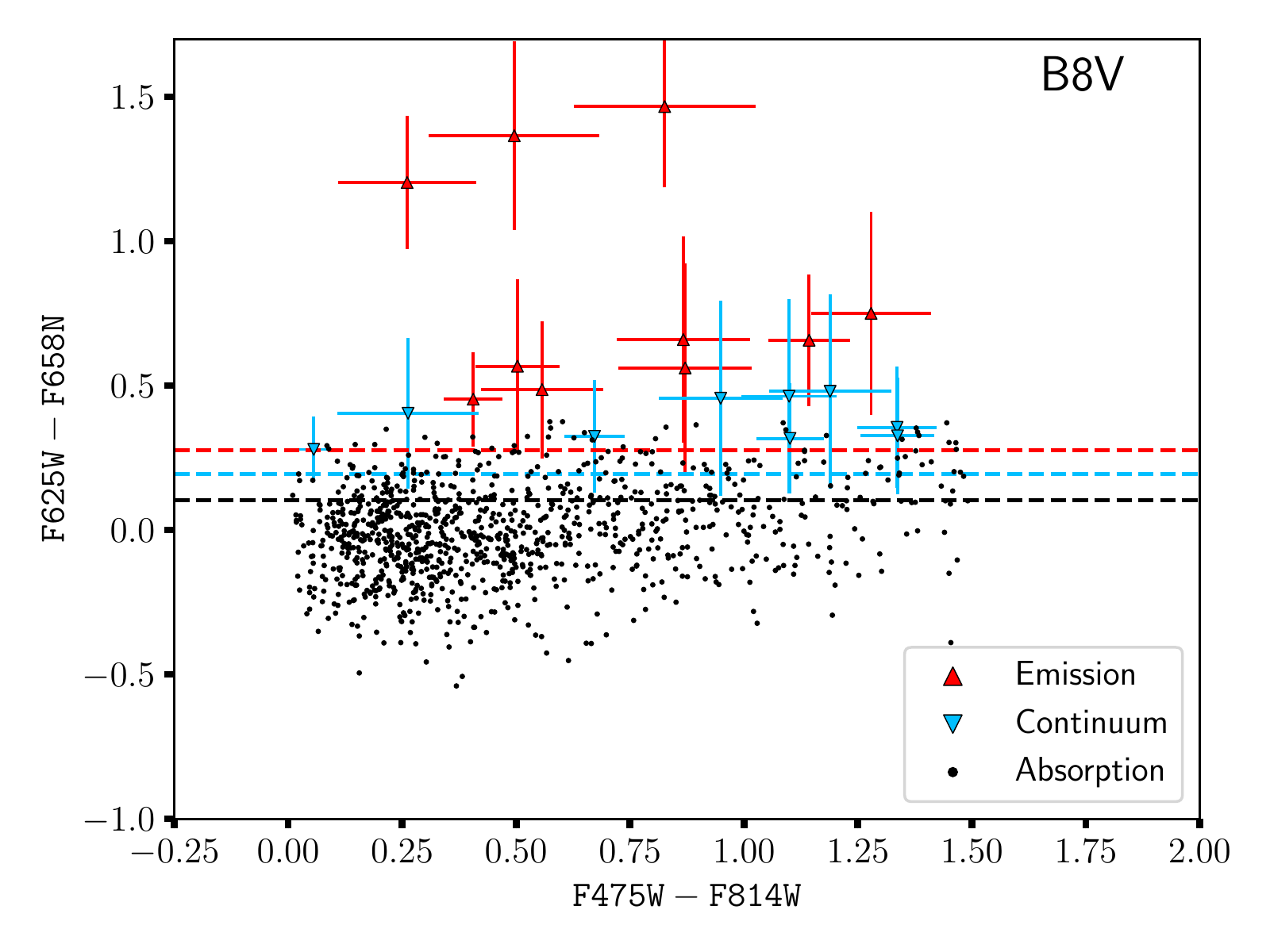}
\caption{A representative 2-CD for our B8V spectral type sources, from 1 epoch of a pointing.  The description of the data presentation is the same as in Figure \ref{2ColorB0V}.}
\label{2ColorB8V}
\end{figure}

Our identification of classical Be stars were made using 2-color diagrams (2-CDs) developed for each of our spectral sub-types (e.g. Figures \ref{2ColorB0V}, \ref{2ColorB1V}, \ref{2ColorB3V}, \ref{2ColorB5V}, and \ref{2ColorB8V}) set by the 50th-percentile T$_{eff}$ reported by the \texttt{BEAST}.  The boundaries of different spectral sub-type assignments are simply the midpoint T$_{eff}$ between sub-types listed in Table \ref{tbl:kurucz}.  Because the T$_{eff}$ values are inferred from broadband photometric observations by the \texttt{BEAST}, we expect our spectroscopic sub-type assignments could differ by 1-2 sub-types.  However, because we bin the spectroscopic classifications in Section 3.3, minor mis-classifications do not impact our overall results.  We also explored effects of including $1\sigma_{err}$ in our 50th-percentile T$_{eff}$ source catalog cuts, and found that this only changed our final fractional Be ratios $\pm$1\%.

The dotted lines in each 2-CD demarcate the expected values for pure photospheric absorption (black), filled-in absorption (blue), and pure emission (red) profiles from our synthetic spectra.  We adopted conservative limits for identifying sources as classical Be stars, requiring their 
(\texttt{F625W}$-$\texttt{F658N}) colors to be redder than the photospheric absorption level (e.g. $[(\texttt{F625W}-\texttt{F658N}) - 3\sqrt{(\texttt{F625W$_{err}$})^2 +
(\texttt{F658N$_{err}$})^2} ]$ $>$ 
photospheric levels).  Stated differently, we conservatively define normal B-type
stars without circumstellar disks as being either (1) located below the dotted black line or (2) include the black line 
within their 3$\sigma$ error interval in Figures \ref{2ColorB0V} - \ref{2ColorB8V}. Such normal B-type stars are depicted by black circles in these figures.

 We further broke down the Be population into two categories; stars whose 3$\sigma$ error interval extended blueward of the pure emission cutoff (potentially weaker disk systems, depicted by blue triangles) and stars whose 3$\sigma$ error interval denoted it unambiguously as a strong disk system (depicted by red triangles). We were unable to identify any correlations between the amplitude of (\texttt{F625W}-\texttt{F658N}) excess and (\texttt{F475}-\texttt{F814W}) colors, indicating that the presence of red optical continuum excesses \citep{Rivinius2013} did not bias us against detecting the strongest disk systems based on our selection criteria.

\subsubsection{Limitations of the 2-CD Technique} \label{limits}

 It is well established that 2-CDs can identify other astrophysical sources that emit at H$\alpha$ besides classical Be stars, such as supergiants, B$[e]$ stars, and Herbig Be stars.  The level of this contamination is expected to be lower when the diagnostic is used in cluster populations (see e.g. \citealt{Wisniewski2006} and references therein).  Moreover, for our survey, as discussed in Section 3.1, we were able to use the value-added characterization of the fundamental stellar properties of our dataset via the \texttt{BEAST} to exclude sources that had surface gravities that indicated they were not main sequence stars.  As such, our 2-CDs should be cleaner from contaminants than other purely photometric investigations.
 
 It is also prudent to consider the limitations of our 2-CD diagnostic given the wealth of H$\alpha$ emission line profile morphologies and variability that classical Be stars have been shown to exhibit.  We explored these limitations both using published literature and compiled spectra of Be stars from the Be Star Spectra (BeSS) database \citep{Neiner2011}.  The ACS/WFC \texttt{F658N} filter has a central wavelength of 6584 \AA, a FWHM of 87.5 \AA, and $>$1\% throughput across $\sim$130 \AA\ (ACS Instrument Handbook v19.0).  As such, this filter is sufficiently broad to capture the full range of H$\alpha$ line profile morphologies for each spectral sub-type of classical Be star.
 
One can readily compare our (\texttt{F625W}-\texttt{F658N}) color selection criteria for Be stars against the range of H$\alpha$ emission line strengths that classical Be stars exhibit, as these criteria are based on the photospheric absorption EWs of each B 
sub-type (Table \ref{tbl:kurucz}). To aid in this process, we show reference (\texttt{F625W}-\texttt{F658N}) colors for filled-in absorption (H$\alpha$ EW = 0) and moderate 
emission strengths ($\mid EW_{emission} \mid$ = $\mid EW_{absorption} \mid$) in Figures \ref{2ColorB0V} - \ref{2ColorB8V}.  We have confirmed that the H$\alpha$ EW of emission line profiles from moderately-sized stable and stochasically variable Be disks are typically well in excess the red dashed line in Figures \ref{2ColorB0V} - \ref{2ColorB8V} (e.g. have emission EWs well in excess of the underlying photospheric absorption line;  \citealt{rivi1998,Neiner2011,draper2014}).  Our selection 
criteria, for example, would recover $\sim$87\% of the 76 star sample of Be stars spanning a range of spectral sub-types, based on single to few epoch H$\alpha$ EWs presented in \citet{slet92,men94,S2014}. We confirmed this is also true for many Be shell stars \citep{rivi2006}, despite the fact that 
their H$\alpha$ profiles are depressed by narrow central absorption components \citep{Neiner2011, draper2014}.  An exception is the subset of shell stars with particularly weak disks, where the net H$\alpha$ EW becomes dominated by the underlying photospheric absorption line.  Even classical Be stars experiencing an active disk dissipation episode often retain H$\alpha$ EWs above our detection criteria, until the end stages of their disk loss (see e.g. the cases of 60 Cyg and $\pi$ Aqr, \citealt{wisniewski2010}).  By contrast, the early stages of disk re-building episodes as well as the weakest Be disks are often characterized by H$\alpha$ EWs at or below filled-in photospheric absorption levels (EW = 0; dashed blue line in Figures \ref{2ColorB0V} - \ref{2ColorB8V}, e.g. \citealt{wisniewski2010,gru2011,Neiner2011}), and so would likely not be detected by our selection criteria.

\subsection{M31's Total Be Star Population} \label{allBe}
Our methodology identified 552 Be stars out of 8981 (B + Be) stars in epoch 1, and 542 Be stars out of 9098 (B + Be) stars in epoch 2. The resultant Be fraction (\# Be / (\# normal B + \# Be)) for each pointing is shown in Table \ref{BeFractionTable}, and is broken down as a function of spectral sub-type in Table \ref{tbl:specproperties}. We note that pointings 1-3, obtained from the WFC3 camera (see Section \ref{sec:ObsDes}), are excluded from these statistical analyses for the remainder of the paper, as the filter utilized (\texttt{F658N}) only sampled the wing of the H$\alpha$, therefore biasing us against robustly detecting weaker disk signals.  We provide comprehensive information about each normal B-type star and classical Be star in the catalog in Table \ref{tbl:detailproperties}.

\begin{figure}[ht]
\includegraphics[width=\columnwidth]{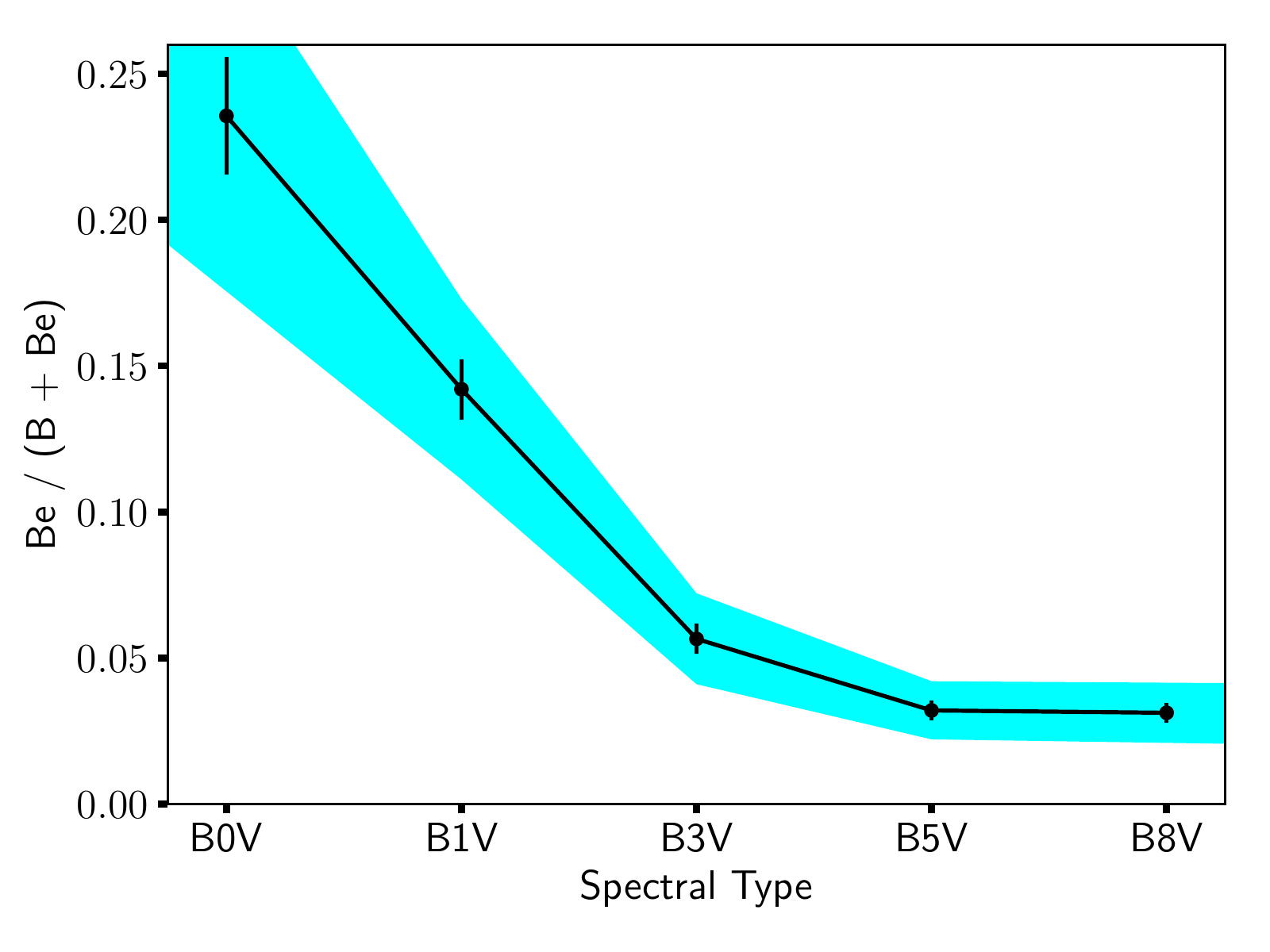}
\caption{The \# Be / (\# normal B + \# Be) fraction as a function of spectral type, as compiled from Table \ref{tbl:specproperties}, is shown.  Note that the black error bars represent the 1$\sigma$ interval and the blue shading represents the 3$\sigma$ interval.}
\label{Fraction}
\end{figure}

The overall average \# Be / (\# normal B + \# Be) stars (in pointings 4-6) is $\sim$6\%, with no statistically 
significant difference between epoch 1 (6.15\% $\pm 0.26\%$) and epoch 2 (5.96\% $\pm0.25\%$).
These overall Be fractions are strongly impacted by the larger number of later spectral type sources we detect, as seen in Table \ref{tbl:specproperties} and Figure \ref{Fraction}. In particular, Figure \ref{Fraction} 
illustrates the clear trend of the Be fraction as a function of spectral type.  The Be fraction is largest for B0-type stars and progressively decreases for later sub-types.

We caution that our detection rates as a function of spectral type in Table \ref{tbl:specproperties} and Figure \ref{Fraction} are likely subject to several biases. The total main sequence B-type star population (i.e. column 4 in Table \ref{tbl:specproperties}) exhibits incompleteness based on expectations from an initial mass function (IMF). Assuming no incompleteness at the B0 sub-class and adopting a Kroupa IMF \citep{kroupa}, our B-type star population is incomplete by a factor of 1.2x (B1), 1.6x (B3), 2.1x (B5), and 6.3x (B8).  Of course if this incompleteness affects our recovery of Be stars at the same rate it does normal B-type stars, the fractional Be content we nominally derive in Table \ref{tbl:specproperties} and Figure \ref{Fraction} would remain unchanged. We also note that the larger (\texttt{F625W}$-$\texttt{F658N}) photometric errors present for later spectral sub-types  
acts as another factor that biases us against detecting weaker disk systems in later spectral sub-types. Since we do not a priori know the intrinsic prevalence of strong versus weaker disk systems as a function of spectral sub-type, we do not quantify an upper limit to our quoted Be fractions owing to this systematic.

\def\arraystretch{1.9}
\begin{table}[t]
\begin{center}
\caption{Summary of \# Be / (\# Be + \# normal B) Stars}\label{BeFractionTable}
\tablewidth{2pc}
\tablecolumns{3}
\begin{tabular}{ccc}
\\
\tableline\tableline
Pointing & Epoch 1 Be Fraction $\pm1\sigma$ & Epoch 2 Be Fraction $\pm1\sigma$\\
\tableline

4 & 5.39\%$\pm0.42\%$ & 5.69\%$\pm0.43\%$\\
5 & 5.25\%$\pm0.40\%$ & 5.45\%$\pm0.40\%$\\
6 & 7.76\%$\pm0.48\%$ & 6.76\%$\pm0.46\%$\\
Overall & 6.15\%$\pm0.26\%$ & 5.96\%$\pm0.25\%$\\

\tableline
\end{tabular}
\tablecomments{This table summarizes the fraction of Be stars (\# Be / (\# Be + \# normal B)) stars detected in each epoch for pointings 4-6, along 
with -$\sigma$ uncertainties. Note that pointings 1-3 are excluded from the overall fractions compiled in this table.}
\end{center}
\end{table}
\def\arraystretch{1.0}

\def\arraystretch{1.9}
\begin{table*}[htp]
\begin{center}
\caption{Summary of \# Be / (\# Be + \# normal B) stars as a function of spectral type }\label{tbl:specproperties}
\tablewidth{2pc}
\tablecolumns{5}
\begin{tabular}{ccccc}
\\
\tableline\tableline
Epoch & Spectral Type & \# Be Stars & (\# normal B + \# Be Stars) & Be Fraction $\pm 1\sigma$\\
\tableline

1 & B0V & 106 & 450 & 23.56\%$\pm2.00\%$\\
1 & B1V & 163 & 1148 & 14.20\%$\pm1.03\%$\\
1 & B3V & 112 & 1981 & 5.65\%$\pm0.52\%$\\
1 & B5V & 91 & 2840 & 3.20\%$\pm0.33\%$\\
1 & B8V & 80 & 2562 & 3.12\%$\pm0.34\%$\\
1 & Overall & 552 & 8981 & 6.15\%$\pm0.26\%$\\

2 & B0V & 107 & 448 & 23.88\%$\pm2.01\%$\\
2 & B1V & 159 & 1152 & 13.80\%$\pm1.01\%$\\
2 & B3V & 119 & 1965 & 6.06\%$\pm0.54\%$\\
2 & B5V & 68 & 2895 & 2.35\%$\pm0.28\%$\\
2 & B8V & 89 & 2638 & 3.37\%$\pm0.35\%$\\
2 & Overall & 542 & 9098 & 5.96\%$\pm0.25\%$\\

\tableline
\end{tabular}
\tablecomments{Be Star fractions for epochs 1 and 2, using pointings 4-6 combined.}
\end{center}
\label{BeFraction}
\end{table*}
\def\arraystretch{1.0}

\subsection{Variability in M31's Be Star Population}

\begin{figure*}[ht]
\includegraphics[width=\textwidth]{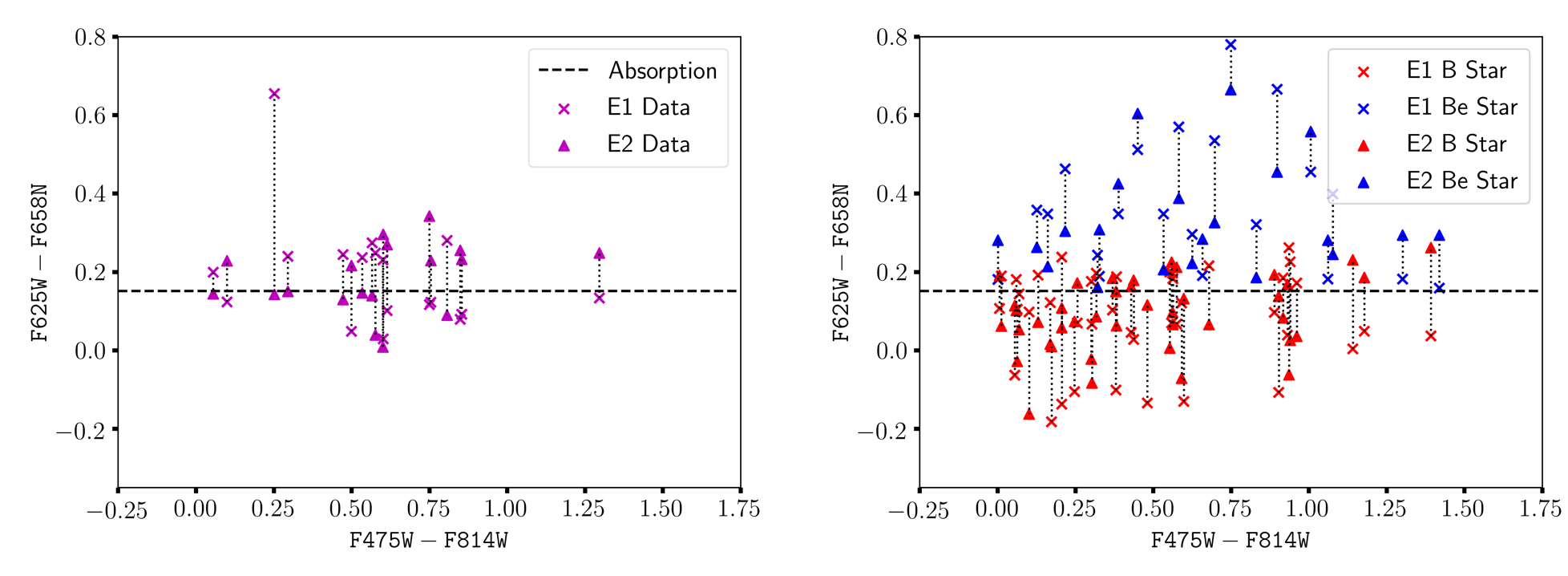}
\caption{A plot of the variability of B0V stars in our dataset. The left hand panel shows how stars that exhibited disk-loss or disk-renewal events (transient Be stars) evolved between the first and second epoch. The right hand plot depicts the variability that stable B and stable Be stars exhibited between the two epochs of observations.} 
\label{VariabilityFig}
\end{figure*}

The wide (year-long) time baseline of our 2 epochs of observations allow us to assess the statistical prevalence of disk-loss and disk-renewal episodes, albeit the poor temporal sampling limits our ability to constrain the detailed time-scale of these events.  Because of the circumstellar disk's H$\alpha$-emitting properties, large changes in the size of these disks will cause sources to significantly move in 2-CDs. We classified a Be star as exhibiting a disk-loss or disk-renewal event if it was determined to be a Be star in one epoch and its (\texttt{F625W}$-$\texttt{F658N}) color was at or bluer than the expected value for a normal B-type star for the other, and vice-versa. We show the evolution of such disk-loss and disk-renewal events in 2-CD space for B0V-type stars in Figure \ref{VariabilityFig}. Following \citet{McSwain2009}, we define the percentage, $p$, of transient Be stars, as the ratio of Be stars that exhibit either disk-loss or disk-renewal events, $n_{Betrans}$, relative to the total number of Be stars, $n_{Betotal}$:

\begin{equation}
p = \frac{n_{Betrans}}{n_{Betotal}}
\end{equation}

We also define the error in this percentage to be:

\begin{equation}
\sigma_{p} = \sqrt{\frac{p(1-p)}{n_{Betotal}}}
\end{equation}
$\sim$13\% of our sample exhibited a disk-loss event while $\sim$10\% of our sample exhibited a disk-renewal event (Table \ref{Betransience}).  $\sim$77\% of our sample maintained their status of having a disk at some level through the 1-year baseline of our data. As seen in Figure \ref{VariabilityFig} for our B0-type stars, such non-transient systems still exhibited evidence of variability in their H$\alpha$ emission. Note that the total \# of Be stars reported in both epochs in Table \ref{Betransience} is lower than the number reported in epoch \# 1 or \# 2 in Table \ref{tbl:specproperties} as some sources had too low SNR to be meet our source catalog requirements in one of two epochs.

\begin{table*}[t]
\begin{center}
\caption{Disk-loss and Disk-renewal episodes}
\label{Betransience}
\tablewidth{2pc}
\tablecolumns{6}
\begin{tabular}{cccccc}
\\
\tableline\tableline
Class & \# Maintained Disk & \# Disk-Loss & \# Disk Renewal & Total \# Be stars in both epochs & p \\
\tableline
B0V & 80 & 9 & 9 & 98 & 18 $\pm$4\% \\
B1V & 121 & 14 & 11 & 146 & 17 $\pm$3\% \\
B3V & 79 & 12 & 15 & 106 & 25 $\pm$4\% \\
B5V & 40 & 13 & 6 & 59 & 32 $\pm$6\% \\
B8V & 32 & 9 & 2 & 43 & 26 $\pm$7\% \\
Total & 352 & 57 & 43 & 452 & 22 $\pm$2\% \\
\tableline
\end{tabular}
\tablecomments{A summary of the number of sources (both cluster and field) that gained or lost a disk through the 1-year duration of our observations.  The percentage of transient systems exhibiting disk-loss or disk-renewal, p, is also given.  
}
\end{center}
\end{table*}

\subsection{Cluster Versus Field Be Stars} \label{clusterdefinition}
Ascertaining whether our source catalog of stars belong to distinct clusters or a general field population
is of interest as the uniform age of cluster members can allow one to assess how the Be phenomenon evolves with a star's fractional main sequence lifetime.  We matched our source catalog against clusters identified by the citizen science-based Andromeda Project and compiled in \citet{Johnson2015}.  Specifically, we required our sources to be within the quoted aperture radius for each cluster, \texttt{R$_{ap}$}, as compiled in  \citet{Johnson2015}.  

Of the 8981 B + Be stars we detected in epoch 1, we classified 420 cluster stars in 85 unique clusters and 8561 as field stars.  We computed a corresponding  fractional Be content of 16.19$\pm$1.80\% in cluster environments and a much smaller fractional Be content of 5.65$\pm$0.25\% in field environments (Table \ref{tbl:cluster}).  Our results from epoch 2 are consistent with those found in epoch 1 to within errors.  In epoch 2, of the 9098 B + Be stars, we classified 417 as cluster stars in 86 unique clusters and 8681 as field stars, and computed a fractional Be content of 14.39$\pm$1.72\% in cluster environments and a much smaller fractional Be content of 5.55$\pm$0.24\% in field environments (Table \ref{tbl:cluster}). Be stars are located in 23 unique clusters in epoch 1 and 27 unique clusters in epoch 2. We caution that our photometry is likely incomplete or of worse quality in the densest cluster environments, where source contamination from blends could be more detrimental.

\begin{figure*}[htp]
\includegraphics[width=\textwidth]{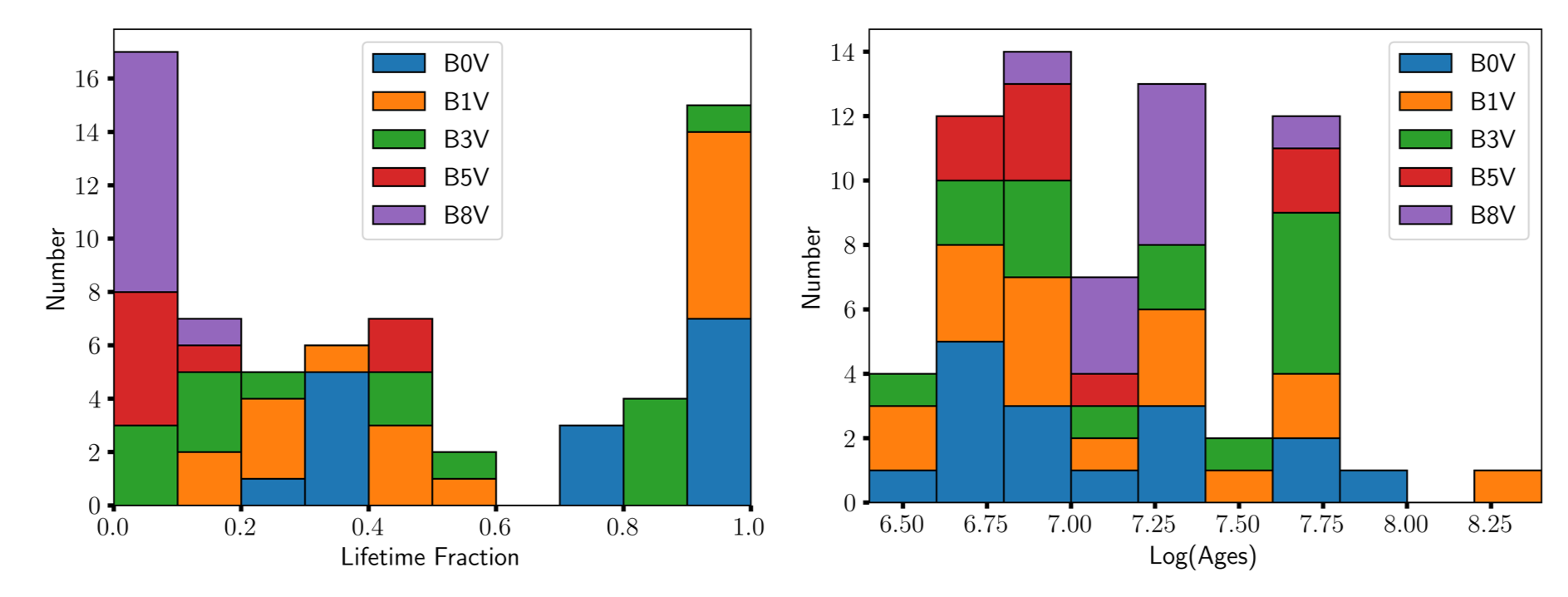}
\caption{The ages of our epoch 1 Be stars located in cluster environments. The left panel shows the distribution of cluster Be stars as a function of their fractional main sequence lifetime, defined as the ratio of a star's age to its main sequence lifetime age, whereas the right hand panel shows the distribution as a function of cluster age. Two cluster stars were omitted from this figure as they lacked robust pre-existing age estimates.} 
\label{ClusterAgeHist}
\end{figure*}

The distribution of ages for our Be stars located in clusters, adopted from the cluster ages determined and tabulated by \citet{Johnson2016}, is shown in Figure \ref{ClusterAgeHist}. We also computed the fractional main sequence lifetime for each cluster star, defined as the ratio of a star's age to its main sequence lifetime. This was done by assigning a mass to each inferred spectral sub-type, taken from Table 1 of \citet{Silaj_2014}, and adopting main sequence lifetimes for Solar metalicity stars, assuming $\Omega$ / $\Omega_{crit,initial}$ = 0.8, from interpolating Table 2 of \citet{Georgy2013}. The resultant distribution of our cluster sample as a function of fractional main sequence lifetime is shown in Figure \ref{ClusterAgeHist}.  We remark that this methodology initially assigned 12 of our sources with fractional main sequence lifetimes exceeding 1, likely due to improper main sequence lifetime assignment related to the 1-2 spectral sub-type uncertainty of our sample (Section \ref{identify}), and were re-assigned fractional main sequence lifetimes of 1.  We further discuss and interpret trends in these data in Section \ref{agediscussion}.

\def\arraystretch{1.8}
\begin{table}[htp]
\begin{center}
\caption{Prevalence of Be stars in cluster versus field environments}\label{tbl:cluster}
\tablewidth{2pc}
\tablecolumns{5}
\begin{tabular}{ccccc}
\\
\tableline\tableline
Epoch & Location & \# Be stars & \# normal B stars & Be fraction $\pm 1\sigma$ \\
\tableline
1 & Field & 484 & 8077 & 5.65\%$\pm0.25\%$\\
1 & Cluster & 68 & 352 & 16.19\%$\pm1.80\%$\\
2 & Field & 482 & 8199 & 5.55\%$\pm0.24\%$\\
2 & Cluster & 60 & 357 & 14.39\%$\pm1.72\%$\\

\tableline
\end{tabular}
\tablecomments{The fractional Be content (\# Be / (\# normal B + \# Be)) for objects identified as likely belonging to cluster and field environments is tabulated for both epochs of our observations.}
\end{center}
\end{table}
\def\arraystretch{1.0}

\section{Discussion} \label{sec:Discussion}

\subsection{Variability}
A remarkable aspect of the Be phenomenon is dramatic episodes of complete disk-loss and disk-regeneration.  Long-term photometric, spectroscopic, and polarimetric studies have suggested that complete dissipation of some disks occurs over time-scales as long as one to several years (e.g. \citealt{wisniewski2010,car2012}), whereas partial disk-growth episodes imprint a variety of clear observational signatures over timescales of days to months to years \citep{rivi1998,gru2011,draper2014,Bartz2017,labadie2018}.  Both observational and theoretical studies have suggested that the disk dissipation phase can last for $\sim$2x longer than the typical time-scale for disk build-up phases \citep{h2012,vieira2017,labadie2018}.  Since each observational bandpass/diagnostic probes different physical regions within Be disks \citep{h2012,Rivinius2013}, it is important to test this different disk growth versus dissipation time-scale prediction of the viscous decretion disk model \citep{h2012} with a variety of observational techniques.  

These events plausibly represent epochs in which the amplitude or efficacy of the disk driving mechanisms change.  Diagnosing both the time-scale and frequency of such events therefore seems likely to elucidate the fundamental mechanism(s) driving the phenomenon.  Early studies of the frequency of  
transient events were limited, based only on small samples of Galactic clusters (e.g. 45 Be stars distributed across 7 clusters; \citealt{McSwain2008,McSwain2009}), revealing a wide dispersion in the observed rate of transients spanning 
a wide range of time-scales (0$\pm$18\% - 75$\pm$22\%, with a composite mean of 51 $\pm$7\%; \citealt{McSwain2009}). \citet{granada2018} characterized the IR color variability and location in IR CMD and 2-CD space of a sample of Be stars in open clusters to infer that $\sim$9-15\% of their sample of Be stars had a dissipating or small disk. Larger volume surveys have quantified that long-term variations (LTVs) in the $R$-band occur over many years in 37\% (81/217) of Be stars \citep{Bartz2017}.  However, LTVs are not classified on the strict basis of complete dissipation of an observational signature of a disk or the regeneration of a disk from a disk-less state, and include partial dissipation and build-up events.  

Our survey provides a significant statistical boost to previous studies of the frequency of transient events, i.e. a 10x larger sample size than \citet{McSwain2009}. Our sample includes both cluster and field stars, which differs from earlier studies that focused solely on cluster populations (e.g. \citealt{McSwain2009}). We found 22 $\pm$ 2\% (100/452 cluster and field Be stars; Table \ref{Betransience}) of Be stars experienced a disk-loss or disk-renewal episode within 1 year, which is lower than the composite mean of 51 $\pm$7\% (23/45 Be stars) reported by the smaller \citet{McSwain2009} study, although these transient rates do technically agree within 3$\sigma$ of the quoted errors. The rate of disk-loss episodes that we observe across our entire sample ($\sim$13\%; Table \ref{Betransience}) is similar to the fraction of dissipating or small disks inferred from the IR photometric study of \citet{granada2018}. We do caution that Be stars have been found to vary over periods of many years, and our one-year survey may not have captured the full transition period for many of these stars.  Indeed, Figure \ref{VariabilityFig} illustrates that many of our non-transient sources exhibit variability, some of which could be revealed to be the onset of transient events given higher photometric accuracy and/or additional epochs of observations.

We note that the compilation of transient Be fractions from \citet{McSwain2008,McSwain2009} were computed from 4 years (NGC 3766) and 2 years (Collinder 272, Jogg 16, IC 2581, NGC 3293, NGC 4755, NGC 6231, NGC 6664) of observations, so it is not clear how much the different durations of these surveys affect these statistical comparisons.  If we make the simple assumption that, to first order, the Be transient fraction over several years scales linearly by a rate set by the duration of observations, the resultant yearly transient Be fractions are remarkably consistent.  Namely, the 22 $\pm$ 2\% yr$^{-1}$ we determine from our present study is consistent with that derived from NGC 3766 (17.3 $\pm$ 3\% yr$^{-1}$; \citealt{McSwain2008}) and the remaining 6 clusters in \citet{McSwain2009} (20.5 $\pm$ 4.5\%).  Since the rate of disk-outburst events has been noted to be more prevalent in early-type stars compared to mid-type Be stars \citep{labadie2018}, it could be interesting to further build up statistics on the \textit{rate} of the transient Be fraction as a function of spectral type, to see if a similar trend emerges.

Finally, we remark that our detection of modestly more disk-loss events (57; Table \ref{Betransience}) compared to disk-renewal events (43; Table \ref{Betransience}) is unexpected given the duration of our sample (1 year) and previous reports that the disk dissipation phase can last for $\sim$2x the typical time-scale for disk build-up phases \citep{vieira2017,labadie2018}.  If disk dissipation is a slow process, we would expect to preferentially detect the more rapid disk renewal events, which is the opposite of what we observe. We caution that our adoption of conservative 2-CD criteria for the presence of a disk, requiring a 3$\sigma$ elevation above H$\alpha$ photospheric levels, could be introducing an observational bias in our data that mis-classifies disks decreasing in mass as disk-loss events, and conversely underestimates small-scale disk-renewal events. 

\subsection{Age Dependence of the Be Phenomenon}\label{agediscussion}

We explored how Be stars identified in cluster environments depended on cluster age and fractional main sequence lifetimes in Section \ref{clusterdefinition}.  Interpreting trends in these data first requires a discussion of the known limitations of our dataset.  Fundamentally, our analysis is enabled by characterization of the fundamental stellar parameters of our dataset by use of the \texttt{BEAST} Bayesian toolset.  However, as noted in Section \ref{identify}, we do expect our spectral sub-typing to have uncertainties of 1-2 sub-types.  We see likely byproducts of these uncertainties in Figure \ref{ClusterAgeHist}.  For example, the presence of early sub-type Be stars in clusters $>$30 Myr old could imply these clusters have experienced multiple episodes of star formation and/or be a reflection that the spectral sub-types assigned to this population are incorrect by 1-2 sub-types. Similarly, we noted in Section \ref{clusterdefinition} that we re-assigned 12 cluster stars to have a fractional main sequence lifetime of 1 (from $>$ 1), which was a likely byproduct of small mismatches in our spectral sub-typing. 

With these caveats in mind, we do see several interesting trends appear in Figure \ref{ClusterAgeHist}.  First, we observe a clear population of Be stars at early fractional main sequence lifetimes.  This supports the idea that a subset of classical Be stars must emerge on the ZAMS as rapid rotators, as suggested by \citet{wisniewski2007}.  Although the enhancement in the number of early-type Be stars observed at late fractional main sequence lifetimes could support the idea that the Be phenomenon is enhanced with evolutionary age \citep{fabregat2000,martayan2010}, we caution that we do not have sufficient statistics with our existing M31 dataset to uniformly sample how the frequency of early type Be stars changes as a function of fractional main sequence lifetime.

\subsection{Be Fraction and Metallicity}

\begin{figure}[t]
\includegraphics[width=\columnwidth]{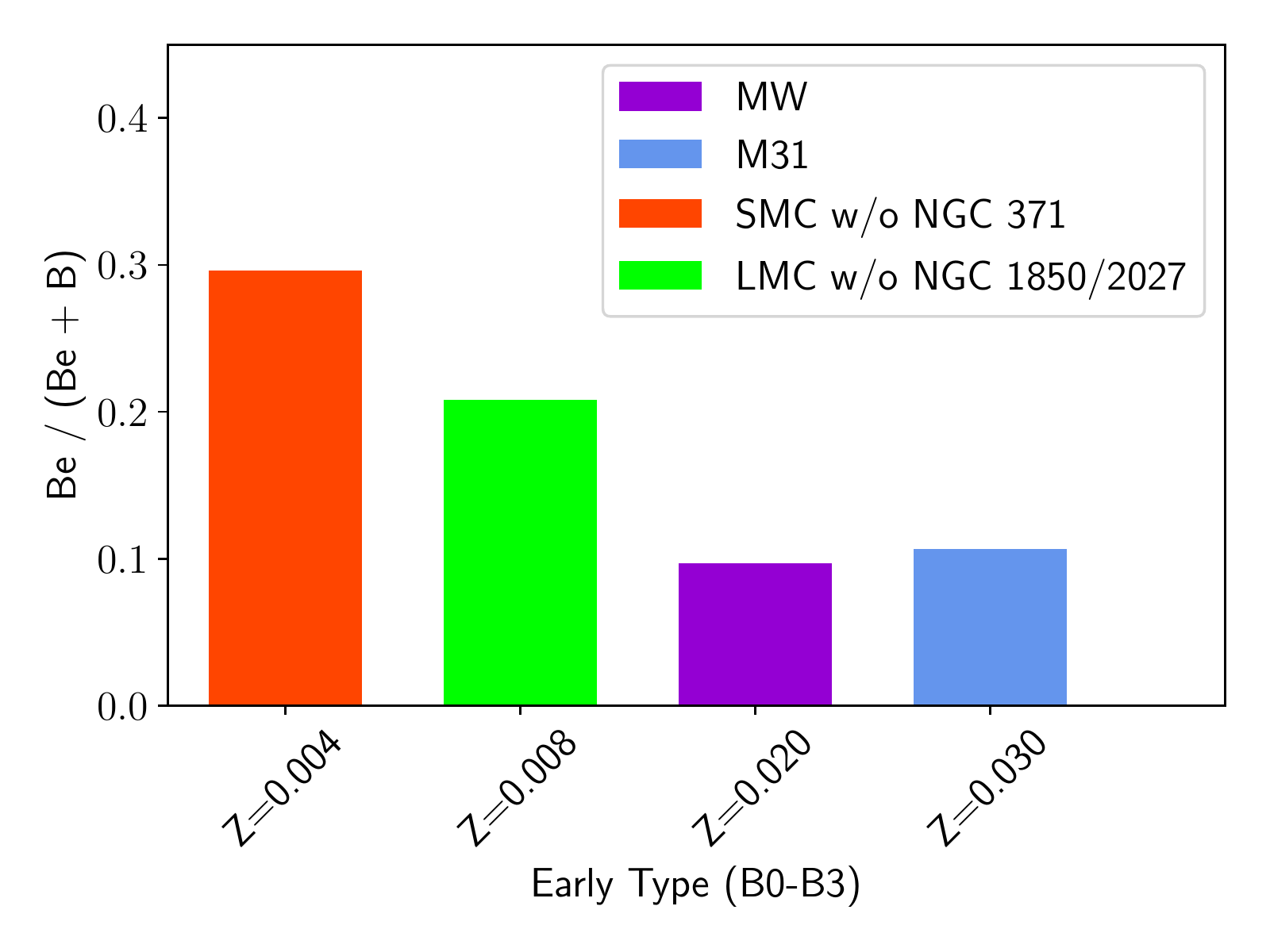}
\caption{The fractional Be content of early-type (B0-B3) stars is shown as a
function of metallicity.  We adopted metallicities (Z) of 0.030 (1.5x solar) for M31, Z = 0.020 for the Milky Way, Z = 0.008 for the LMC, and Z = 0.004 for the SMC.  The LMC and SMC data are taken from Table 6 of \citet{Wisniewski2006}, which includes data from \citet{Maeder1999}, and \citet{Keller99}}; we choose to omit several clusters with abnormally large samples sizes that would otherwise bias our sample averages (e.g. NGC 371 in the SMC; NGC 1850 and NGC 2027 in the LMC).  We did attempt to include additional data (e.g. \citealt{Martayan2007}), however discrepancies in the available online data prevented uniform inclusion of these data.  Our MW data was obtained from \citet{McSwain2005} and \citet{Wisniewski2006}; we assigned spectral sub-types to these data based on criteria compiled in \citet{martayan2010}.
\label{MetallityComparisons}
\end{figure}

Many studies of Galactic, LMC, and SMC Be stars have found that the fractional Be content of an environment is related inversely to its metallicity  \citep{Maeder1999,Wisniewski2006,Martayan2006,martayan2010,iq2013}.  The reported amplitude of this trend varies substantially in the literature, likely due to differences in completeness, filter systems, disparate stellar ages, and non-uniform ranges of spectral types amongst the datasets.  This trend has been suggested to arise from differences in the average rotation rates of stars in these environments, with low metallicity environments having a larger population of rapid rotators (e.g. \citealt{mae2000,mae2001,mar07}).  Since the present day metallicity of M31 is approximately 1.5x solar \citep{Clayton2015}, our dataset allows us to explore this trend on an extended metallicity range.  

The spectral-type dependence of the fractional Be content has 
been extensively studied in the Galaxy, LMC, and SMC (see e.g. \citealt{Zorec1997,McSwain2005,Wisniewski2006,mat2008,martayan2010,Rivinius2013}). Although the decrease in percentage of Be stars with spectral type that we observed in M31 (Figure \ref{Fraction}) is generally consistent with that reported in the literature in other environments, significant attention must be paid to factors such as sample size, completeness, and the methodology used to quantify Be stars in each study.

The number of single epoch Be stars we identified for statistical analysis in our M31 sample ($\sim$550 Be stars; Table \ref{tbl:specproperties}), is 1.6x larger than the full Galactic sample used by \citet{McSwain2005}, 2.0x larger than the full Galactic sample used by \citet{mat2008}, 1.7x larger than the slitless SMC sample of \citet{martayan2010}, 1.2x larger than the photometric SMC sample of \citet{Wisniewski2006}, and 1.2x larger than the photometric LMC sample of \citet{Wisniewski2006}. However, despite our large sample size, the total number of main-sequence B-type stars (\# normal B + \# Be) that we find as a function of spectral sub-type (Table \ref{tbl:specproperties}) is flat through the mid to late B-type stars. This deviation from that expected based on a nominal initial mass function distribution indicates our M31 sample is incomplete at mid- to late- spectral sub-types, similar to that reported for all SMC and LMC studies (see e.g. \citealt{Wisniewski2006,martayan2010}).  

If we limit the detailed comparison of our M31 results with the literature to our earliest spectral sub-types, where completeness is less of an issue, we do find some interesting results.  Figure \ref{MetallityComparisons} depicts the fractional Be content for early-type (B0-B3) stars as a function of metallicity, using Galactic Be star data from \citet{McSwain2005}, LMC and SMC Be star data from \citet{Wisniewski2006}, \citet{Maeder1999}, and \citet{Keller99}, and M31 Be star data from the present study.  The fractional Be content we observe in M31 is lower than that observed in the SMC or LMC, in agreement with metallicity trends previously noted in the literature \citep{Maeder1999,Wisniewski2006,Martayan2006,martayan2010,iq2013}.  While Be stars are more prevalent in the SMC than M31 for the earliest sub-types by a factor of 2.8x, this ratio is still lower than the 3-5x enhancement between the SMC and Milky Way reported by \citet{martayan2010}. By contrast, our data indicate no clear change in the frequency of early-type Be stars in M31 and Milky Way, despite their different metallicity. Since the current star formation rate of M31 is lower than that of the Milky Way \citep{yin2009,barb2010,alexia2015}, it is possible that our results are a reflection that the Be phenomenon is enhanced by evolutionary age as stars progress through their main sequence evolutionary lifetimes.

\subsection{Comparison with Previous M31 Studies} 
The SPLASH survey \citep{dorman2012} conducted a Keck II/DEIMOS optical spectroscopic survey of $\sim$5000 isolated stars in M31, including regions observed by the PHAT survey.  The SPLASH survey took into account the amount of crowding present during targeting, to reduce the role of nearby neighbor contamination, and targeted sources brighter than \texttt{F814W} = 22 or \texttt{F475W} = 24 \citep{Prichard2017}. The survey identified a total of 146 main sequence Be stars that overlapped with the PHAT footprint, and determined a net fraction Be content (\# Be / (\# normal B + \# Be)) of $\sim12\%$ \citep{Prichard2017}.

\begin{figure}[htp]
\includegraphics[width=\columnwidth]{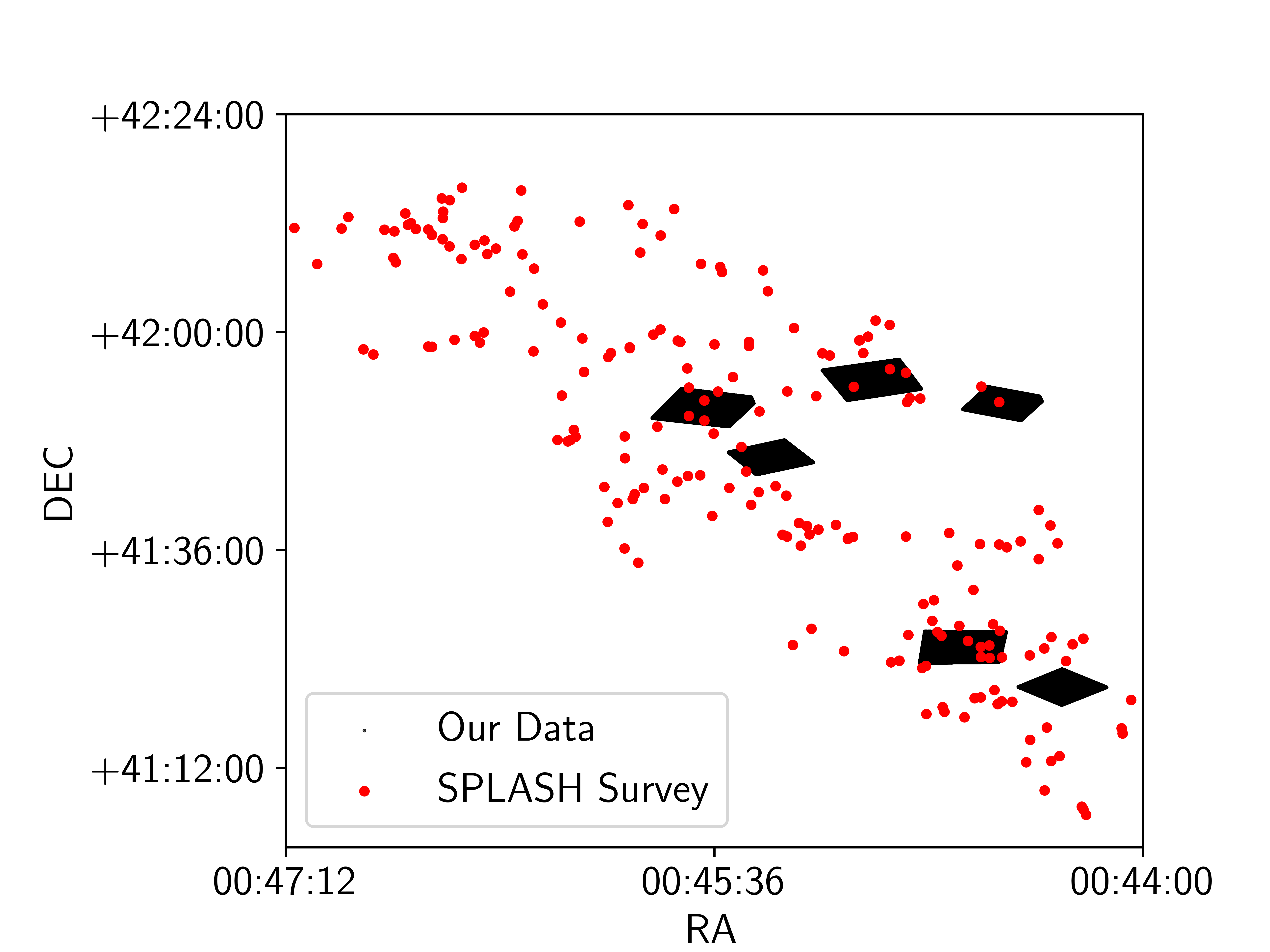}
\caption{A plot of the RA and DEC of stars listed in the SPLASH H$\alpha$ survey as emission stars with stars in our pointings in black.}
\label{keckfoot}
\end{figure}

We identified a total of 14 emission line stars in the SPLASH H$\alpha$ emission survey that overlapped with our HST footprint. These include 10 stars classified as Be 
stars, 2 classified as RHeB, 1 classified as AGB, and 1 classified as T-MS (Figure \ref{keckfoot}).  
SPLASH utilized 0$\farcs$8 wide slitlets on Keck/DEIMOS.  To investigate the potential role of source confusion and contamination in these spectra, we computed the number of sources in our HST footprint that fell within the Keck/DEIMOS aperture size for each of the 10 Be stars identified by \citet{Prichard2017} that fell within our FOV.  We found 24 HST-detected sources in total were present within the Keck aperture for these 10 Be stars, and provide a summary of each of the HST-detected sources that fell within these Keck/DEIMOS apertures in Table \ref{deimos}. Most of these HST-detected sources had \texttt{BEAST}-derived stellar parameters that were inconsistent with them being main sequence B-type stars; 7 HST-detected sources had T$_{eff}$ more consistent with O-type stars and 14 HST-detected sources had Log(g) values more consistent with post-main sequence giants. As seen in Table \ref{deimos}, each SPLASH Be star was matched with one source having a bright \texttt{F475W} magnitude, along with several less luminous stars. It is therefore possible that the H$\alpha$ EWs reported by \citet{Prichard2017} have slightly elevated (or suppressed) signal from these fainter neighbors, that could lead to spurious detections of Be stars in the survey.
We note that none of the stars that we identified as Be stars in the analysis of our HST data were matched with Be stars identified by the SPLASH survey. 

Given the clear dependence of the Be phenomenon as a function of spectral sub-type that we see in our M31 sample (Figure \ref{Fraction} and Table \ref{tbl:specproperties}) and the shallower depth of source brightnesses probed by the SPLASH survey, we caution that the global (spanning all spectral sub-types) M31 Be fraction of $\sim$12\% cited by \citet{Prichard2017} is strongly biased by overweighting contributions from the earliest B sub-types.  

On a broader context, we noted in Section 3.5 that our HST survey also likely suffered from source confusion, especially in the densest cluster regions of M31. If the effects of source confusion serves to produce a greater frequency of false-positive detections of classical Be stars, as seems possible from our overlapping HST coverage of the SPLASH H$\alpha$ footprint, this would imply that the Be fractions we derive in our study could be similarly affected by false-positive detections, especially in the densest stellar environments.

\begin{table*}
\scriptsize\addtolength{\tabcolsep}{-0.15pt}
\begin{center}
\caption{Be Star Match Data}\label{deimos}
\tablewidth{2pc}
\tablecolumns{10}
\begin{tabular}{ccccccccccc}
\\
\tableline\tableline
\textbf{SPLASH Survey} Index & Pointing & RA & DEC & \texttt{F475W} & \texttt{F625W}$-$\texttt{F658N} & Exceeds 3$\sigma$ Expected & \texttt{F475W}$-$\texttt{F814W} & Temperature (K) & \textbf{Log(g)$_{50}$} & \textbf{Log(g)$_{84}$}  \\
\tableline
20276 & 6 & 00:44:36.4 & +41:25:19.5 & 19.37 & 0.22 & Unknown & 0.16 & 37103 & 3.62 & 3.72 \\
20276 & 6 & 00:44:36.4 & +41:25:19.2 & 23.80 & 0.32 & True & 2.04 & 23061 & 3.09 & 3.58 \\
\tableline
21673 & 6 & 00:44:34.4 & +41:24:06.4 & 20.62 & 0.15 & Unknown & -0.07 & 42266 & 4.18 & 4.28 \\
21673 & 6 & 00:44:34.4 & +41:24:06.1 & 24.82 & -0.06 & False & 0.62 & 9096 & 3.52 & 3.65 \\
21673 & 6 & 00:44:34.4 & +41:24:06.2 & 26.11 & -0.14 & False & 1.62 & 13293 & 4.11 & 4.32 \\
21673 & 6 & 00:44:34.4 & +41:24:06.3 & 25.70 & 0.39 & False & 1.06 & 12244 & 3.99 & 4.21 \\
\tableline
21767 & 6 & 00:44:34.4 & +41:25:28.8 & 19.75 & 0.97 & Unknown & 0.46 & 49538 & 3.99 & 4.10 \\
21767 & 6 & 00:44:34.4 & +41:25:28.8 & 19.75 & 0.86 & Unknown & 0.46 & 49538 & 3.99 & 4.10 \\
\tableline
22397 & 6 & 00:44:45.2 & +41:26:33.8 & 21.31 & 0.28 & True & 0.53 & 29677 & 3.61 & 3.83 \\
22397 & 6 & 00:44:45.1 & +41:26:34.0 & 25.48 & 0.11 & False & 1.09 & 5385 & 2.69 & 2.89 \\
22397 & 6 & 00:44:45.2 & +41:26:33.6 & 25.05 & -0.43 & False & 0.57 & 10796 & 3.74 & 3.95 \\
22397 & 6 & 00:44:45.2 & +41:26:33.6 & 25.72 & 0.45 & False & 1.52 & 4823 & 2.48 & 2.68 \\
\tableline
22675 & 4 & 00:45:38.3 & +41:50:15.7 & 22.97 & 0.17 & False & 0.74 & 31634 & 4.11 & 4.24 \\
22675 & 4 & 00:45:38.3 & +41:50:15.7 & 26.56 & -0.08 & False & 1.81 & 5360 & 2.69 & 2.89 \\
22675 & 4 & 00:45:38.3 & +41:50:15.9 & 26.44 & 0.28 & False & 1.75 & 4899 & 2.54 & 2.76 \\
\tableline
22760 & 2 & 00:44:32.2 & +41:52:17.6 & 22.09 & 0.19 & False & 0.78 & 23175 & 3.41 & 3.60 \\
22760 & 2 & 00:44:32.2 & +41:52:17.3 & 26.29 & --- & False & 1.83 & 4913 & 2.55 & 2.77 \\
\tableline
22776 & 4 & 00:45:38.3 & +41:52:28.5 & 21.03 & 0.14 & Unknown & 0.09 & 34252 & 4.00 & 4.12 \\
22776 & 4 & 00:45:38.3 & +41:52:28.1 & 25.95 & -1.46 & False & 1.81 & 4946 & 2.50 & 2.67 \\
\tableline
22859 & 5 & 00:45:04.8 & +41:53:59.3 & 21.09 & 0.84 & Unknown & 0.60 & 46804 & 4.08 & 4.25 \\
22859 & 5 & 00:45:04.8 & +41:53:59.0 & 26.34 & -0.33 & False & 2.05 & 4717 & 2.42 & 2.60 \\
\tableline
23050 & 5 & 00:44:53.1 & +41:55:32.4 & 20.98 & 0.21 & Unknown & 0.58 & 38616 & 3.88 & 4.08 \\
\tableline
23073 & 5 & 00:44:56.7 & +41:55:54.9 & 22.85 & 0.41 & True & 0.15 & 15174 & 3.59 & 3.70 \\
23073 & 5 & 00:44:56.7 & +41:55:54.9 & 26.24 & 0.00 & False & 1.86 & 5938 & 2.72 & 2.97 \\
\tableline
\end{tabular}
\tablecomments{Shows the relevant data for the stars within the aperture radius of stars in the SPLASH survey. Log(g)$_{x}$ is the x'th percentile of the Log(g) posterior probability distribution and "Exceeds 3$\sigma$ Expected" denotes whether the star meets the emission line criteria  outlined in Section \ref{sec:Analysis}. Some stars in our survey are assigned identical RA/DEC coordinates due source confusion in the PHAT catalog.}
\end{center}
\label{DataCompare}
\end{table*}

\def\arraystretch{1.2}
\begin{table*}
\begin{center}
\scriptsize\addtolength{\tabcolsep}{-2.8pt}
\caption{Source Catalog} \label{tbl:detailproperties}
\tablecolumns{16}
\begin{tabular}{cccccccccccccccc}
\tableline\tableline
Pointing & Epoch & Index & RA & DEC & \texttt{F475W} & \texttt{F625W} & \texttt{F658N} & \texttt{F814W} & Log(T)$_{eff}$ & \textbf{Log(g)$_{50}$} & \textbf{Log(g)$_{84}$} & Cluster & Be Star? & Spec Type & Var?\\
\tableline
1 & 1 & 1907281 & 11.369 & 41.785 & 21.309$\pm{0.003}$ & 21.242$\pm{0.008}$ & 21.284$\pm{0.042}$ & 21.175$\pm{0.004}$ & 4.505 & 3.949 & 4.104 & 0 & False & B0V & BSTAR\\
1 & 1 & 1022497 & 11.371 & 41.76 & 21.733$\pm{0.004}$ & 21.361$\pm{0.008}$ & 21.567$\pm{0.043}$ & 21.248$\pm{0.005}$ & 4.496 & 3.827 & 3.983 & 0 & False & B0V & BSTAR\\
1 & 1 & 1907276 & 11.383 & 41.774 & 21.244$\pm{0.003}$ & 21.170$\pm{0.008}$ & 21.135$\pm{0.032}$ & 21.078$\pm{0.004}$ & 4.396 & 3.562 & 3.781 & 0 & False & B1V & BSTAR\\
... & ... & ... & ... & ... & ... & ... & ... & ... & ... & ... & ... & ... & ... & ... & ... \\
\\
\tableline
\end{tabular}
\tablecomments{A summary of all sources detected in each of our pointings for both epochs, including their index numbers, RA, and Dec as tabulated in \citet{williams2014}.  Note that the full version of this Table is available in the online version of this manuscript.  Photometry from our observations, along with broad-band 
photometry from the PHAT survey as compiled in \citet{williams2014} are provided for each source. All photometry errors are cited at the 1-$\sigma$ level.  The quoted T$_{eff}$ and Log(g) values were computed via 
the \texttt{BEAST}, where Log(g)$_{x}$ is the x'th percentile of the Log(g) posterior probability distribution \citep{gordon2016}.  The \texttt{Cluster} column indicates our best estimate for cluster membership, with 0 denoting a field star and other numbers 
denoting the cluster ID from \citet{Johnson2015}.  The \texttt{Be Star?} column quantifies whether we have determined the star is a classical Be star, using simple ``T(rue)'' and ``F(alse)'' descriptors. The \texttt{Spec Type} 
column provides our best estimate for the spectral type of the star based on its T$_{eff}$.  The \texttt{Var?} column denotes 
whether the star exhibited variability between epochs. The descriptor ``MAIN'' denotes a star that maintained its disk between epochs, ``LOST'' denotes a star that experienced a disk-loss episode, ``GAIN'' denotes a star that experienced a disk-renewal episode, ``BSTAR'' denotes stars that were normal B-type stars in both epochs, 
and ``UNKWN'' is a star that was detected in only one of the epochs.}
\end{center}
\end{table*}
\def\arraystretch{1.0}

\section{Conclusions}
We have presented results from a 12-orbit, 2-epoch HST H$\alpha$ emission line survey of the Andromeda Galaxy that overlapped with the footprint of the HST/PHAT survey, focusing on the classical Be star 
content of the Galaxy.  $\sim$2 million sources were detected in each epoch; after applying criteria to isolate the classical Be star population in this sample, we found: \begin{itemize}
    \item 552 (542) classical Be stars and 8429 (8556) normal B-type stars were identified in epoch \# 1 (epoch \# 2).  The overall fractional Be content (\# Be / (\# normal B + \# Be)) we found was 6.15\% $\pm$0.26\% (5.96\% $\pm$0.25\%) in epoch \# 1 (epoch \# 2).  These overall rates are lower than that reported by the Keck II/DEIMOS-based SPLASH survey ($\sim$12\%, \citealt{Prichard2017}).  We suggest this discrepancy arises from the selection criteria for the SPLASH survey overweighting contributions from the earliest B sub-type stars, which are more likely to exhibit the Be phenomenon, as well as potential source confusion in SPLASH.
    
    \item The fractional Be content decreased with spectral sub-type from $\sim$23.6\% $\pm$2.0\% ($\sim$23.9\% $\pm$2.0\%) for B0-type stars to $\sim$3.1\% $\pm$0.26\% ($\sim$3.4\% $\pm$0.35\%) for B8-type stars in epoch \# 1 (epoch \# 2). We caution that our later-type Be population likely suffers from incompleteness.
    
    \item The fractional Be content we determine for B0-B3 type stars in M31 is lower than that observed for the same subset of spectral type objects in the SMC and LMC. This provides confirmation that the fractional Be content in more metal rich environments (the Milky Way and $\sim$1.5x Solar M31) is lower than that observed in metal poor environments (LMC and SMC). We find that Be stars are more prevalent in the SMC than M31 for the earliest sub-types by a factor of 2.8x; this ratio is lower than the 3-5x enhancement between the SMC and Milky Way reported by \citet{martayan2010}. 
    
    \item We observe a clear population of Be stars in cluster environments at early fractional main sequence lifetimes. This supports the idea that a subset of classical Be stars emerge onto the ZAMS as rapid rotators. Since M31 has a lower star formation rate compared to the Milky Way, the lack of an observed change in the frequency of early-type Be stars between these two galaxies, despite their 1.5x metallicity difference, may reflect that the Be phenomenon is also enhanced by evolutionary age.
    
    \item 22\% $\pm$ 2\% of our sample exhibited evidence of either a  disk-loss or disk-regeneration episode in the $\sim$1 year interval between our two epochs of observations.  If we assume that the Be transient fraction over time-scales of several years scales linearly by a rate set by the duration of observations, the 22 $\pm$ 2\% yr$^{-1}$ transient rate we observe in M31 is consistent with that derived from NGC 3766 (17.3 $\pm$ 3\% yr$^{-1}$; \citealt{McSwain2008}) and the remaining 6 clusters in \citet{McSwain2009} (20.5 $\pm$ 4.5\%).   We observed a similar number of disk-loss events (57) as disk-renewal events (43), which was unexpected as disk dissipation 
    time-scales can be $\sim$2x the typical time-scales for disk build-up phases \citep{vieira2017,labadie2018}.
    
\end{itemize}

\acknowledgements
JW thanks Karen Bjorkman for stimulating discussions of the classical Be phenomenon and this paper.  This work was supported by STScI grants for GO-13857, and by NSF-AST 1412110.  This work has made use of the BeSS database, operated at LESIA, Observatoire de Meudon, France: http://basebe.obspm.fr

\software{\texttt{DOLPHOT} \citep{dolphot2000}, \texttt{pysynphot} \citep{pysynphot2013}, \texttt{BEAST} \citep{gordon2016}, \texttt{matplotlib} \citep{matplotlib}, \texttt{scipy} \citep{scipy}}

\nocite{*}

\pagebreak

\bibliographystyle{yahapj}
\bibliography{main}

\appendix
\section{Focus-Induced Systematics} \label{oops}
During our basic assessment of the quality of our pipeline photometry products, we noted a $-0.1$ magnitude offset between epochs in our \texttt{F625W} photometry of pointings 2, 3, 5, and 6.  No such offset were seen in our \texttt{F658N} data, or between epochs for pointings 1 and 4.  Moreover, we observed clear variations in the shape of the PSF from epoch-to-epoch difference images.  This suggests that the photometric offsets we observe are due to HST focus variations that are not otherwise accounted for in our photometry. It is well known that HST experiences "breathing" over the course of an orbit, causing the distance between the primary and secondary mirror (the "despace") to vary on the order of $\pm$5 $\mu$m from nominal. This causes corresponding variations in the PSF; when the secondary mirror position is off-nominal (i.e., has a nonzero despace value), light is proportionally less concentrated in the core of the PSF and more concentrated in the first and second Airy rings.  \citet{Dressel2012} demonstrated that for WFC3/UVIS F420M the fraction of encircled energy within a 0$\farcs$15 aperture decreases by over 20\% going from nominal to +5 micron despace.

In most scenarios, any PSF variation induced by focus is not the dominant systematic in HST photometry; most long exposures ($>$ $\sim$2000 seconds) and co-adds of multiple short exposures are likely to sample the short-term focus variation adequately, and if one uses empirically determined PSFs for photometry or TinyTim \citep{Krist2011} PSFs at correct focus values, then the effects of focus will most likely be accounted for to within photometric error. As our photometry uses a single model PSF at nominal focus (see e.g. \citealt{williams2014}), we are sensitive to focus-induced systematics. It is also worth noting that the focus model shows moderate deviation from measured focus values, and that there are other factors affecting focus that are not accounted for by the model (van Gorkom 2016, private communication); however, for our purposes we are satisfied that the focus deviations predicted by the model account for the photometric offsets we observe.

We measured the focus for each individual exposure using the HST focus tool\footnote{http://focustool.stsci.edu/cgi-bin/control.py}, which implements the focus model described in \citet{Cox2012}.  For our epoch 1 data, we found a despace value close to nominal for pointings 2, 3, 5, and 6, and between 3-4 microns for pointings 1 and 4. For our epoch 2 data, all pointings had despace values between 3-4 microns.  To correct our data for this constant offset and ensure the same relative flux calibration is present across our entire dataset, we therefore subtracted 0.1 magnitudes from pointings 1 and 4 in epoch 1 and pointings 1-6 in epoch 2.

\end{document}